\documentclass[10pt]{article}

\usepackage[margin=1.2in]{geometry}
\usepackage{framed,multirow}

\usepackage{latexsym, amssymb, enumerate, amsmath}
\usepackage{defs}
\usepackage{fonts}
\usepackage{dsfont}
\usepackage{pifont}
\usepackage{epsfig}
\usepackage{epstopdf}
\usepackage{booktabs}
\usepackage{lineno}
\usepackage{xfrac}
\usepackage{siunitx}
\usepackage[normalem]{ulem}

\usepackage{subfig} 
\usepackage{enumitem}
\usepackage{algorithm}
\usepackage{algpseudocode}

\usepackage{xargs}                      
\usepackage{upgreek}

\graphicspath{{./figures/}}

\usepackage{url}

\title{A fast implicit solver for semiconductor models in one space dimension\thanks
	{This manuscript has been authored, in part, by UT-Battelle, LLC, under Contract No. DE-AC0500OR22725 with the U.S. Department of Energy. The United States Government retains and the publisher, by accepting the article for publication, acknowledges that the United States Government retains a non-exclusive, paid-up, irrevocable, world-wide license to publish or reproduce the published form of this manuscript, or allow others to do so, for the United States Government purposes. The Department of Energy will provide public access to these results of federally sponsored research in accordance with the DOE Public Access Plan (\texttt{http://energy.gov/downloads/doe-public-access-plan}).}}

\date{\today}
\author{M. Paul Laiu%
\thanks{Computational Mathematics Group,
				Computer Science and Mathematics Division,
				Oak Ridge National Laboratory,
				Oak Ridge, TN 37831 USA, (\texttt{laiump@ornl.gov}).
}       
\and
Zheng Chen%
\thanks{
Department of Mathematics,
University of Massachusetts Dartmouth,
285 Old Westport Road, 
Dartmouth, MA 02747 USA, (\texttt{zchen2@umassd.edu}).
}
\and
        Cory D. Hauck%
\thanks{Computational Mathematics Group,
				Computer Science and Mathematics Division,
				Oak Ridge National Laboratory,
				Oak Ridge, TN 37831 USA, (\texttt{hauckc@ornl.gov}).
This author's research was sponsored by the Office of Advanced Scientific
Computing Research and performed at the Oak Ridge National Laboratory,
which is managed by UT-Battelle, LLC under Contract No. DE-AC05-00OR22725.}
}

\begin{document}

\maketitle

\begin{abstract}
Several different approaches are proposed for solving fully implicit discretizations of a simplified Boltzmann-Poisson system with a linear relaxation-type collision kernel.  This system models the evolution of free electrons in semiconductor devices under a low-density assumption.
At each implicit time step, the discretized system is formulated as a fixed-point problem, which can then be solved with a variety of methods.  A key algorithmic component in all the approaches considered here is a recently developed sweeping algorithm for Vlasov-Poisson systems.
A synthetic acceleration scheme has been implemented to accelerate the convergence of iterative solvers by using the solution to a drift-diffusion equation as a preconditioner. 
The performance of four iterative solvers and their accelerated variants has been compared on problems modeling semiconductor devices with various electron mean-free-path. 
\end{abstract}



\section{Introduction}
\label{sec:intro}

The Boltzmann-Poisson system is considered an accurate kinetic model of electron transport in semiconductor devices \cite{Markovich-Ringhofer-Schmeiser-1990}.
This system describes the evolution of an electron distribution function using a semi-classical Boltzmann kinetic equation and generates a self-consistent electric field by coupling the Boltzmann equation to a Poisson equation that is driven by the electron density.
Numerical simulation of the Boltzmann-Poisson system is known to be difficult for several reasons, including the nonlinear coupling between equations, the nonlinear collision operator that describes electron-electron and electron-background interactions, and the dimension of the computational domain. Indeed, simulating a three-dimensional device requires the solution to a six-dimensional Boltzmann equation.

Under a low-density assumption, electron-electron interactions become negligible and electrons can be treated as classical particles interacting with a material background.  In such cases, the nonlinear collision operator can be replaced by a linear, relaxation time approximation \cite{Markovich-Ringhofer-Schmeiser-1990,ferry1991physics,Cercignani2001iv} when the steady-state solution to the Boltzmann-Poisson system is of interest.
In the case that the electron transport is through channels that are parallel to the electric field, the semiconductor device is effectively one-dimensional \cite{Cercignani2001iv,Markovich-Ringhofer-Schmeiser-1990} and can therefore be simulated using the approximated Boltzmann-Poisson system in one space dimension.

In addition to traditional Direct Simulation Monte Carlo methods \cite{jacoboni1989monte}, many deterministic numerical schemes have been developed for solving the Boltzmann-Poisson system and its simplified variants.
The deterministic schemes considered in previous works discretize the position-velocity phase space of the Boltzmann equation and its simplified variants using the weighted essentially non-oscillatory (WENO) finite difference method \cite{Carrillo2000iza,Carrillo2003,Carrillo2006}, the discontinuous Galerkin (DG) method \cite{Cheng2009hm,Cheng2008,Cheng2011ew}, and the spectral-difference method \cite{ringhofer2003mixed}.
These schemes either consider a steady-state Boltzmann-Poisson system or use explicit time-stepping schemes to capture transient behavior.
To guarantee stability, explicit time-stepping schemes usually require the size of the time steps to be proportional to the mean-free-path of particles in the system.
Such a restriction can be computational prohibitive for highly collisional problems, where the mean-free-path is small.
An explicit asymptotic preserving scheme was introduced to address this issue in \cite{Jin-Pareschi-2000}, where the stability is guaranteed under a parabolic time-step restriction (independent of the mean-free-path) for highly collisional problems.
Later, an implicit-explicit (IMEX) asymptotic preserving scheme was developed in \cite{dimarco2014implicit} to relax the parabolic time-step restriction.  
However, these and similar approaches do not allow for large-scale variations in the mean-free-path that are common in multiscale problems.

In this paper, we consider a fully implicit numerical scheme for solving the simplified Boltzmann-Poisson system under the low density assumption in one space dimension.
The fully implicit time-stepping method allows for larger time steps that are independent of the mean-free-path, regardless of the collisionality of the problem.
Such stability comes at the cost of solving large, possibly ill-conditioned, linear and nonlinear algebraic equations.  Hence efficient numerical solvers are needed to update the numerical solution at each time step.
In \cite{Garrett-Hauck-2018}, a fast, fully implicit solver was proposed for the nonlinear Vlasov-Poisson system, which is the collisionless variant of the simplified Boltzmann-Poisson system considered in this paper.
At each time step, this solver applies a special decomposition of the phase space to allow for the use of the sweeping technique that are commonly used to accelerate the solution of radiation transport problems \cite{Adams-Larsen-2002,Larsen-Morel-2010,Lewis-Miller-1984}.
To utilize the fast solver \cite{Garrett-Hauck-2018} in the collisional case, we consider the scattering term as a source and formulate the simplified Boltzmann-Poisson system as a nonlinear fixed-point problem at each implicit time step. 
At each fixed-point iteration, a collisionless problem with source is solved and the electron distribution that solves the collisionless problem is used to update the collision term, which becomes the source at the next fixed-point iteration.
The fixed-point problem reaches a solution when the collisionless problem gives an electron distribution that is consistent with the collision term.

Two types of fixed-point formulations for the simplified Boltzmann-Poisson systems are considered and compared in this paper.
The main difference between the two formulations is in the treatment of the scattering source in the relaxation operator: in the first case, both the electric field and the scattering source are lagged; in the second, only the electric field is lagged.  
As a result, problems in the first formulation are solved in a single iteration loop, while the ones in the second formulation require two nested loops.   
We apply various iterative solvers on these problems and compare their performance. 
To solve problems in the first formulation, we consider  Picard iteration  (see, e.g., \cite[Section~I.8]{hairer1993solving}) and also Anderson acceleration \cite{Anderson-1965,Walker-Ni-2011}.
For problems in the second formulation, we solve the nonlinear outer loop via Anderson acceleration and solve the linear inner loop using either Picard iteration or the generalized minimal residual method (GMRES) \cite{Saad-Schultz-1986}.  
We do not apply Picard iteration on the outer loop since preliminary numerical results suggest that this approach is not competitive.

We also consider accelerated/preconditioned variants of the solvers described above, based on the idea of synthetic acceleration (SA) \cite{lebedev1964iterativeKP,kopp1963synthetic,Adams-Larsen-2002,Lewis-Miller-1984}, an approach that accelerates convergence of iterative solvers by applying a correction term in between each iteration.
This correction term is obtained by solving coarse, cheap, or low-order approximate equations to the error equation of the base iterative solver.
Thus, many of the SA schemes can be viewed as two-level multigrid algorithms \cite{Larsen1991} or preconditioned iterative solvers \cite{Adams-Larsen-2002,derstine1985use,Bruss-Morel-Ragusa-2014}.
For neutron transport problems, the correction terms can be computed by solving the transport equation on a coarse mesh \cite{lorence1989s,Bruss-Morel-Ragusa-2014} or solving a diffusion equation \cite{Alcouffe-1976,Alcouffe-1977,Adams-Larsen-2002} that is a low-order approximation to the transport equation near the collision limit where the mean-free-path is small.
In this paper, we compute the correction term by solving a drift-diffusion equation that approximates the simplified Boltzmann-Poisson system in the highly collisional, low-field regime.

The various strategies are tested on a one-dimensional silicon $n^+$--$n$--$n^+$ diode problem \cite{Carrillo2000iza,Cercignani2000domain,Cheng2009hm,Cercignani2001iv,Hu-2014}.  
The algebraic equations to be solved are derived via a backward Euler discretization in time and a discontinuous Galerkin (DG) discretization in the position-velocity phase space.
The low-order time discretization is chosen primary to simplify the presentation; however the DG discretization is important for capturing the drift-diffusion limit.
While other discretizations are possible, our focus here is on the efficiency of the solver strategy.
Thus for various levels of collisionality, the computation time and iteration count of each fixed-point formulation and each iterative solver are compared.
These results provide a guideline on the selection of iterative solvers for problems with different material profiles.

The remainder of the paper is organized as follows.  In Section \ref{sec:prelim},
the simplified Boltzmann-Poisson system, the drift-diffusion equation, and the time, space, and velocity discretization for solving them are described. 
Section \ref{sec:techniques} provides details of fixed-point formulations for the discretized equations as well as the iterative solvers for these fixed-point problems.
In Section \ref{sec:num_results}, the implementation details and numerical results for the various iterative solvers are reported for the $n^+$--$n$--$n^+$ diode problem. 
Conclusions and discussion are given in Section~\ref{sec:conclusion}.

\section{Preliminaries}
\label{sec:prelim}

\subsection{Semiconductor models}
\label{subsec:semiconductor_eqn}

We consider the kinetic model
\begin{subequations}
\label{eq:semiconductor_model}
\begin{equation}
\label{eq:semiconductor_eqn}
\p_t f + v \p_x f + \frac{q_e}{m} E \p_v f = \omega (\rho M_{\Theta} - f)\:,
\end{equation}
\begin{equation}
\label{eq:Poisson_eqn}
E = \p_{x} \Phi\:,\quad \p_{x}^2 \Phi = \frac{q_e}{\varepsilon_p} (\rho - D)\:.
\end{equation}
\end{subequations}
Here \eqref{eq:semiconductor_eqn} describes the evolution of the electron distribution $f = f(t,x,v)$, which is a function of position $x \in [0,L]$, velocity $v \in \bbR$, and time $t \geq 0$; the electric field $E = E(t,x)$ in \eqref{eq:Poisson_eqn} is the spatial gradient%
\footnote{There is a sign difference between the electric field defined in \eqref{eq:semiconductor_model} and the usual physics definition convention. We make this choice to match the definition used in the fast sweeping algorithm in \cite{Garrett-Hauck-2018} for solving Vlasov-Poisson systems.  In particular, the sign choice in \eqref{eq:semiconductor_model} implies that sign of $E$ determines the direction of flow in the velocity variable.  The sweeping algorithm is used extensively in this paper. See Section~\ref{subsubsec:fast_sweeping} for details.}
of a potential $\Phi = \Phi(t,x)$ that satisfies a Poisson equation with a source due to the balance between a given doping profile $D = D(x)$ and the particle (electron) concentration $\rho = \rho(t,x) =\int_{\bbR} f(t,x,v) \text{d}v$.
The constants $q_e$, $m$, and $\varepsilon_p$ denote, respectively, the magnitude of the electron charge, the effective electron mass, and the electric permittivity of the material. 
The collision frequency $\omega = \omega(x)$ takes the form $\omega = \frac{q_e}{m \mu}$ with electron mobility $\mu = \mu(x)$, and the absolute Maxwellian 
\begin{equation}
M_{\Theta}(v) := (2\pi \Theta)^{-\frac{1}{2}} e^{-{v^2}/{2\Theta}}\:,
\end{equation}
where the background temperature $\Theta := \frac{k_\textup{B}}{m} T$ with $k_\textup{B}$ the Boltzmann constant and $T$ the lattice temperature.
For the detail derivations of this model, we refer to the reader to \cite{Markovich-Ringhofer-Schmeiser-1990,selberherr2012simulation}.

\subsubsection{Scaled semiconductor models}
\label{subsubsec:scalings}

Since the qualitative behavior of solutions to \eqref{eq:semiconductor_model} largely depends on the scales of the system, we introduce non-dimensional variables $\hat{t} = \frac{t}{t_0}$, $\hat{x} = \frac{x}{x_0}$, $\hat{v} = \frac{v}{v_0}$ and express \eqref{eq:semiconductor_model} in terms of the scaled variables
\begin{equation}
\hat{f}(\hat{t},\hat{x},\hat{v}) = \frac{f(t,x,v)}{f_0}\:,
\quad
\hat{\omega}(\hat{x}) = \frac{\omega(x)}{\omega_0}\:,
\quad 
\hat{D}(\hat{x}) = \frac{D(x)}{D_0}\:,
\quad
\hat{\rho}(\hat{t},\hat{x}) = \frac{\rho(t,x)}{\rho_0}\:.
\end{equation}
Let $\Phi_0$ be a nominal value of the potential $\Phi$ and let $\Delta\Phi(t,x):=\Phi(t,x)-\Phi_0$.
Because the solution of \eqref{eq:semiconductor_model} is independent of $\Phi_0$, we consider the scaled potential $\hat{\Phi} = \Phi_0 + \frac{\Delta \Phi}{[\Phi]}$.
By assuming $\hat{\rho}=\int_\bbR \hat{f} \text{d} \hat{v}$, the scaled system takes the form
\begin{subequations}
\label{eq:final_semiconductor_model}
\begin{align}
\label{eq:final_semiconductor_eqn}
\delta \p_{ {t}}  {f} +&  {v}  \p_{ {x}}  {f} + \beta^2   {E}  \p_{ {v}}  {f} = \frac{ {\omega}}{\epsilon} ( {\rho} {M}_{\alpha^2} -  {f})\:,\\
\label{eq:final_Poisson_eqn}
  {E} &= \p_{ {x}}  {\Phi}\:,\quad 
\p_{ {x}}^2  {\Phi} = \frac{\gamma^2}{\beta^2} (\zeta {\rho} -  {D})\:,
\end{align}
\end{subequations}
where the hats on the variables dropped for simplicity.
In \eqref{eq:final_semiconductor_model}, the kinetic Strouhal and Knudsen numbers \cite{Golse-Levermore-2005,saint2009hydrodynamic} are given by $\delta = \frac{x_0}{v_0 t_0}$ and $\epsilon = \frac{v_0 }{x_0\omega_0}$, respectively, and the ratios are defined as $\zeta:=\frac{\rho_0}{D_0}$, $\alpha:= \frac{\Theta^{\sfrac{1}{2}}}{v_0}$, $\beta:= \frac{B_0}{v_0}$, and $\gamma:= \frac{C_0}{v_0}$ with
\begin{equation}\label{eq:velocities}
\Theta^{\sfrac{1}{2}}=\left(\frac{k_\textup{B}}{m} T\right)^{\sfrac{1}{2}} \:,\quad B_0:= \left(\frac{q_e [\Phi]}{m}\right)^{\sfrac{1}{2}}\:,\quand 
C_0:= x_0 \omega_{\textup{pe}} := x_0 \left(\frac{q_e^2 D_0 }{\varepsilon_p m}\right)^{\sfrac{1}{2}}\:.
\end{equation}
Here $\Theta^{\sfrac{1}{2}}$ is the thermal velocity, $B_0$ is the ballistic velocity, and in plasma physics, $\omega_{\textup{pe}}$ is known as the plasma frequency \cite{boyd2003physics,hazeltine2004framework}.

\subsubsection{The drift-diffusion limit}
\label{subsubsec:low_field_scalings_DD}

To describe semiconductors with different characteristics, there exist various scaling of the semiconductor model \eqref{eq:final_semiconductor_model}, such as the low-field scaling \cite{Markovich-Ringhofer-Schmeiser-1990,Poupaud-1991}, the high-field scaling \cite{stichel1994asymptotic,poupaud1992runaway,Cercignani2001iv}, and the ballistic scaling \cite{Carrillo2000iza}.
In this paper, we use the solution of a drift-diffusion equation as a preconditioner to accelerate the solution procedure for \eqref{eq:final_semiconductor_model} in the low-field, highly collisional regime. 
While this preconditioner is expected to work well only in this regime, the discretizations of \eqref{eq:final_semiconductor_model} and the formulation of solvers for the resulting algebraic equations do not rely on any particular scaling.

The low-field scaling of \eqref{eq:final_semiconductor_model} assumes that the ratio $\beta$ is an $\mathcal{O}(1)$ quantity and that the ratio $\zeta=1$, i.e., the scaling of the particle concentration $\rho$ and the doping profile $D$ is identical.
Under these assumptions, when $\epsilon$ is small, i.e., when the electron mean-free-path $\lambda:=\frac{v_0}{\omega_0}$ is much smaller than the spatial scale $x_0$, the collision term on the right-hand side of \eqref{eq:final_semiconductor_eqn} is much larger than the drift term $\beta^2 E \p_v f$; it thus becomes the dominant term.
In this situation, it is necessary to choose $\delta \approx \epsilon$ in order to observe the nontrivial dynamics in the long time scale, in which case \eqref{eq:final_semiconductor_model} can be approximated by a drift-diffusion-Poisson model.

It is shown in \cite{Poupaud-1991} that when the potential $\Phi$ is known and sufficiently smooth, a standard drift-diffusion model can be derived from \eqref{eq:final_semiconductor_eqn} in the low-field, collision limit $(\delta\approx\epsilon\to0)$ by expanding the distribution function $f$ via the Hilbert expansion as
\begin{equation}
f(t,x,v) = \varrho(t,x) {M}_{\alpha^2}(v) + \mathcal{O}(\epsilon)\:.
\end{equation}
This result has been extended in \cite{abdallah2004diffusion} and \cite{Masmoudi-Tayeb-2007} respectively to the one-dimensional and multi-dimensional Boltzmann-Poisson systems with a self-consistent potential via Poisson coupling as in \eqref{eq:final_Poisson_eqn}.
The resulting drift-diffusion-Poisson model takes the form
\begin{subequations}
	\label{eqn:DDP}
	\begin{align}
	\label{eq:drift_diffusion_eqn}
	\xi \p_{t} {\varrho} -& \p_{{x}} {\left( \omega^{-1} \p_{{x}} \varrho \right)} + \beta^2 \p_{{x}} { \left( \omega^{-1} E \varrho \right) } = 0\:,\\
	\label{eq:Poisson_eqn2}
		&{E} = \p_{{x}} {\Phi}\:,\quad 
		\p_{{x}}^2 {\Phi} = \frac{\gamma^2}{\beta^2} ({\varrho} - {D})\:,
	\end{align}
\end{subequations}
where \eqref{eq:drift_diffusion_eqn} is a drift-diffusion equation coupled with the Poisson system \eqref{eq:Poisson_eqn2}, and $\xi>0$ is the ratio between $\delta$ and $\epsilon$, i.e., $\delta = \xi \epsilon$.
We refer to this low-field, collision limit as the ``drift-diffusion limit".

In this paper, we use the solution to the drift-diffusion equation as a preconditioner for solving the scaled semiconductor kinetic equation \eqref{eq:final_semiconductor_eqn}. 
The numerical method we used to solve the drift-diffusion equation is discussed in Section~\ref{subsec:drift-diffusion}, and the drift-diffusion preconditioning approach is introduced in Section~\ref{subsec:acce_precond}.

\subsection{Solving the kinetic equation}
\label{subsec:kinetic}

In this paper, implicit time discretization of the kinetic equation \eqref{eq:final_semiconductor_eqn} is considered.
In the simplified case when the electric field $E$ and the particle concentration $\rho$ are known a priori, implicit time discretization of \eqref{eq:final_semiconductor_eqn} leads to linear systems that can be solved efficiently via the fast sweeping algorithm proposed in \cite{Garrett-Hauck-2018}.
The implicit time discretization, the position-velocity phase space discretization, and the fast sweeping approach for solving \eqref{eq:final_semiconductor_eqn} in the simplified case are discussed in Sections~\ref{subsubsec:time}, \ref{subsubsec:phase_space}, and \ref{subsubsec:fast_sweeping}, respectively.
Based on the method presented in this section, we propose several iterative solvers in Section~\ref{sec:techniques} for solving \eqref{eq:final_semiconductor_eqn} in the self-consistent case that $E$ and $\rho$ are coupled via the Poisson equation \eqref{eq:final_Poisson_eqn}.

\subsubsection{Time discretization}
\label{subsubsec:time}
In the temporal domain $[0, \, t_{\textup{final}}]$, we apply a uniform discretization with time step size $\dt$ and denote $f^{n}\approx f(t^n,\cdot,\cdot)$, where $t^n=n\dt$.
To simplify the presentation, we consider the backward Euler scheme in this paper.
Although this scheme is only first-order accurate, it can be used as a building block for higher-order implicit schemes, such as the  singly diagonally implicit Runge-Kutta method (SDIRK) \cite{hairer1993solving,hairer1996solving}.
Applying the backward Euler scheme to \eqref{eq:final_semiconductor_eqn} leads to
\begin{equation}
\label{eq:backward_euler}
v \p_x f^{n+1} + \beta^2  E^{n+1} \p_v f^{n+1} + \left(\frac{\delta}{\dt}+\frac{\omega}{\epsilon}\right)f^{n+1} = \frac{\omega}{\epsilon}\rho^{n+1} M_{\alpha^2} + \frac{\delta}{\dt}f^{n}\:,
\end{equation}
where $\rho^{n+1} = \int_\bbR f^{n+1}\text{d}v$ and $E^{n+1}$ is coupled via the Poisson equation \eqref{eq:final_Poisson_eqn} with $\rho=\rho^{n+1}$.
For the remainder of Section~\ref{subsec:kinetic}, we assume that $\rho^{n+1}$ and $E^{n+1}$ are known a priori at time $t^n$.
In this simplified case, \eqref{eq:backward_euler} becomes a linear steady-state Vlasov problem with a source:
\begin{equation}
\label{eq:steady_state}
v \p_x f + \eta E \p_v f + \sigma f = q\:,
\end{equation}
where, in an abuse of notation, $f = f^{n+1}(x,v)$ denotes the steady-state unknown, $E = E^{n+1}(x)$ denotes a given electric field, $\eta$ and $\sigma$ are positive constants, and $q=q(x,v)$ denotes a general source.
This simplified steady-state problem \eqref{eq:steady_state} can be solved efficiently using the fast sweeping algorithm proposed in \cite{Garrett-Hauck-2018}.
We give the details of this algorithm in Section~\ref{subsubsec:fast_sweeping}.

\subsubsection{Phase space discretization}
\label{subsubsec:phase_space}
For the position-velocity ($x$-$v$) phase space, we truncate the velocity domain from $\bbR$, the entire real line, to a finite interval $[a_v,\,b_v]$. 
The position-velocity computational domain is then $\Dom:=[0,\,L]\times[a_v,\,b_v]$.
Let $\p \Dom$ be the boundary of $\Dom$ and $n(x,v)\in \bbR^2$ be the outward normal at  $(x,v) \in \p \Dom$.   As usual, we decompose the boundary into two disjoint pieces: $\p \Dom = \p \Dom^{\rm{in}} \cup  \p \Dom^{\rm{out}}$ where
\begin{equation}
\p \Dom^{\rm{in}} := \{ (x,v) \colon  (v,E) \cdot n(x,v) \leq 0 \} 
\quand  
\p \Dom^{\rm{out}} := \{ (x,v) \colon  (v,E) \cdot n(x,v) > 0 \}\:.
\end{equation} 
We further decompose $\p \Dom^{\rm{in}}$ into pieces:  $\p \Dom^{\rm{in}} = \p \Dom^{\Lambda} \cup \p \Dom^{\rm{Data}}$, where $\p \Dom^{\Lambda}$ is the portion of the inflow boundary that depends on the interior solution, e.g., periodic or reflecting boundary, and $\p \Dom^{\rm{Data}}$ is the portion upon which data is given.
With these notations, the steady state equation \eqref{eq:steady_state} takes the form
\begin{equation}
\label{eq:steady_state_boundary}
\begin{cases}
v \p_x f + \eta E \p_v f + \sigma f = q &\\
f(x,v) = (\Lambda f) (x,v),  & (x,v) \in \p \Dom^{\Lambda}\\
f(x,v) = w(x,v),  & (x,v)  \in \p \Dom^{\rm{Data}}
\end{cases}
\end{equation}
where $w$ is known and the abstract linear operator $\Lambda$, which maps functions on $\p\Dom^{\text{out}}$ to functions on $\p \Dom^{\Lambda}$, can be used to describe periodic or reflecting boundary conditions.

The computational domain $\Dom$ is discretized into $N_x \times N_v$ rectangular cells of uniform size $\dx\times\dv$.
For $i=1,\dots,N_x$ and $j=1,\dots,N_v$, the cell $C_{i,j}$ is centered at $(x_i,v_j):=((i-\frac{1}{2}) \dx, a_v +(j-\frac{1}{2}) \dv)$.
We denote the set of all cells by $\cT$ and the set of all edges by $\cF$. 
The set $\cF$ is then decomposed into disjoint sets
\begin{equation}
\cF = \cF^{\Lambda} \cup \cF^{\rm{Data}} \cup \cF_x \cup \cF_v \:,
\end{equation}
where $\cF^{\Lambda}$ contains cell edges on the boundary component $\p \Dom^{\Lambda}$, $\cF^{\rm{Data}}$ contains cell edges on the boundary component $\p \Dom^{\rm{Data}}$, and $\cF_x$ and $\cF_v$ contains the remaining cell edges that are perpendicular to the $x$ and $v$ axes, respectively. 
We further decompose $\cF^{\Lambda}=\cF^{\Lambda}_x\cup\cF^{\Lambda}_v$ and $\cF^{\rm{Data}}=\cF^{\rm{Data}}_x\cup\cF^{\rm{Data}}_v$, where the subscripts denote the axis to which the edges are perpendicular.

Let $\cZ = \{g \in L^2 (\bbR \times \bbR) \colon g|_{\Dom^c} = 0 \}$ and $\cZ^h:=\{g^h\in\cZ \colon g^h|_{C} \in \bbP^1(C)\:,\,\forall C\in\cT\}$, where $\bbP^1(C)$ denotes the space of polynomials up to degree one on the cell $C$.
For $g^h\in\cZ^h$, the traces on the two sides of an edge $e\in\cF_x\cup\cF_v$ are defined as 
\begin{equation} 
g^{h,\pm}(x,v) = \lim_{\epsilon\to 0^+} g^h (x\pm\epsilon , v\pm\epsilon)\:,\quad \text{for all } (x,v) \in e\:.
\end{equation}
For these edges, the numerical trace of $g^h$ is defined via upwinding.
Specifically, let $\bar{v}_{e_x}$ denote the value of $v$ at the center of edge $e_x\in\cF_x$, and let $\bar{E}_{e_v}$ denote the value of $E$ at the center of edge $e_v\in\cF_v$, the numerical traces on $e_x$ and $e_v$ are respectively defined as
\begin{equation}\label{eq:numerical_traces}
\hat{g}^h(x,v)=
\begin{cases}
g^{h,-}(x,v)\:, & \bar{v}_{e_x}>0 \:,\\
g^{h,+}(x,v)\:, & \bar{v}_{e_x}<0\:,
\end{cases}
\quad
\hat{g}^h(x,v)=
\begin{cases}
g^{h,-}(x,v)\:, & \bar{E}_{e_v}>0 \:,\\
g^{h,+}(x,v)\:, & \bar{E}_{e_v}<0\:.
\end{cases}
\end{equation}
This definition guarantees a constant upwind direction on each edge.
For a test function $g^h\in\cZ^h$, the traces on a boundary edge 
$e\in\cF^{\Lambda} \cup \cF^{\rm{Data}}$ 
are defined as
\begin{equation}
g^{h,\p}(x,v) = \lim_{\epsilon\to 0^+} g^h (x-\epsilon n_x, v-\epsilon n_v)\:,\quad \text{for all } (x,v) \in e\:,
\end{equation}
where $n_x$ and $n_v$ are the first and second components of the outward normal $n$, respectively.

With these definitions, the discontinuous Galerkin method solves for $f^h\in\cZ^h$ that satisfies
\begin{equation}\label{eq:variational_formulation}
\cA(f^h,g^h)=\cQ(g^h)\:,\quad\forall g^h\in\cZ^h\:,
\end{equation}
with the bilinear operator 
\begin{equation}
\begin{alignedat}{2}
\cA(f^h,g^h):=&\sum_{C\in\cT}\iint_{C} (-\bar{v}_C f^h \p_x g^h - \eta \bar{E}_C f^h \p_v g^h +\sigma f^h g^h )  \dd x \dd v \\
&- \sum_{e\in\cF_x}\bar{v}_e \int_{e} \hat{f}^h (g^{h,+} - g^{h,-}) \dd v
- \sum_{e\in\cF_v}\eta \bar{E}_{e} \int_{e} \hat{f}^h (g^{h,+} - g^{h,-}) \dd x \\
&- \sum_{e\in\cF^{\Lambda}_x}|\bar{v}_e| \int_{e} (\Lambda f^h) g^{h,\p} \dd v
- \sum_{e\in\cF^{\Lambda}_v}\eta|\bar{E}_e| \int_{e} (\Lambda f^h) g^{h,\p} \dd x
\end{alignedat}
\end{equation}
and the source 
\begin{equation}
\cQ(g^h):= \sum_{C\in\cT} \iint_{C} q g^h \dd x \dd v 
+ \sum_{e\in\cF^{\rm{Data}}_x}|\bar{v}_e| \int_{e} w g^{h,\p} \dd v
+ \sum_{e\in\cF^{\rm{Data}}_v}\eta|\bar{E}_e| \int_{e} w g^{h,\p} \dd x\:,
\end{equation}
where $\bar{v}_C$ and $\bar{E}_C$ are the values of $v$ and $E$ at the center of cell $C$, respectively.

In the case of neutral particles ($E=0$), the upwind definition of the numerical traces in \eqref{eq:numerical_traces} allows \eqref{eq:variational_formulation} to be solved with an explicit sweeping procedure that moves through the computational domain in a direction determined by the sign of $v$.
Such sweeping procedures are commonly used for solving radiation transfer problems \cite{Adams-Larsen-2002,Lewis-Miller-1984,Larsen-Morel-2010}.
However, in the case of charged particles, the procedure no longer applies when both $v$ and $E$ are allowed to change sign over the phase space domain.  This is because, in such cases, changes in the upwind direction may create cyclic dependencies in the elements, see, e.g., \cite[Figure~3.1]{Garrett-Hauck-2018}.
To address this challenge, a domain decomposition approach was introduced in \cite{Garrett-Hauck-2018} to break these dependencies. 
We briefly discuss this  approach and the associated sweeping method in the next subsection.

\subsubsection{Domain decomposition and fast sweeping}
\label{subsubsec:fast_sweeping}

The domain decomposition method separates the phase space into subdomains along the line $\{v=0\}$. 
Let $\Dom_1:=[0,\,L]\times(0,\,b_v]$ and $\Dom_2:=[0,\,L]\times[a_v,\,0)$ be the subdomains, and let $\Gamma:=[0,\,L]\times \{0\}$.
We assume that there exists some index $j_0$ such that $C_{i,j_0}\subseteq \Dom_2$ and $C_{i,j_0+1}\subseteq \Dom_1$, and denote the set of cell edges in $\Gamma$ as $\cF_0$.
We further decompose $\cF_0$ into disjoint sets $\cF_0^+$ and $\cF_0^-$ based on the sign of the electric field at the edge center. 
For $g^h\in\cZ^h$, we define
\begin{equation}
g^h_1(x,v)= 
\begin{cases}
g^{h}(x,v)\:, & (x,v)\in \Dom_1\:,\\
0\:, & \text{otherwise}\:,
\end{cases}
\quad
g^h_2(x,v)= 
\begin{cases}
g^{h}(x,v)\:, & (x,v)\in  \Dom_2\:,\\
0\:, & \text{otherwise}\:.
\end{cases}
\end{equation}
The bilinear form in \eqref{eq:variational_formulation} then can be expanded as
\begin{equation}\label{eq:decomposed_bilinear}
\cA(f^h,g^h)=\cA(f^h_1,g^h_1)+\cA(f^h_2,g^h_2)+\cA(f^h_1,g^h_2)+\cA(f^h_2,g^h_1)\:.
\end{equation}
By the definition of $\cA$, it is straightforward to verify that
\begin{subequations}
\begin{align}
\label{eq:coupled_1}
\cA(f^h_1,g^h_2) &= - \sum_{e\in\cF_0^-}\eta |\bar{E}_{e}| \int_{e} \hat{f}^h  g^{h,-} \dd x\:,\\
\label{eq:coupled_2}
\cA(f^h_2,g^h_1) &= - \sum_{e\in\cF_0^+}\eta |\bar{E}_{e}| \int_{e} \hat{f}^h  g^{h,+} \dd x\:,
\end{align}
\end{subequations}
where the edges in $\cF_0^+$ do not appear in \eqref{eq:coupled_1} since $f_1^h$ does not contribute to the numerical traces $\hat{f}^h$ on these edges due to upwinding. Similarly, the edges in $\cF_0^-$ are not included in \eqref{eq:coupled_2}.
We then define for all $\cF_0^{\pm} \subset \Gamma$, the edge values 
\begin{equation}
\hat{f}^{h,*}_1 = \cP_1(f_1^h) :=
\begin{cases}
\hat{f}_1^h\:, & \text{on }\cF_0^-\:,\\
0\:, &  \text{on }\cF_0^+\:,
\end{cases}
\quand
\hat{f}^{h,*}_2 = \cP_2(f_2^h):=
\begin{cases}
\hat{f}_2^h\:, & \text{on }\cF_0^+\:,\\
0\:, &  \text{on }\cF_0^-\:.
\end{cases}
\end{equation}
The system \eqref{eq:variational_formulation} is then equivalent to the coupled system
\begin{subequations}
\label{eq:coupled_sys}
\begin{align}
\label{eq:coupled_sys_1}
\cA(f^h_1,g^h_1) &= \cQ(g^h_1)- \cB(\hat{f}^{h,*}_2,g^h_1)\:,\\
\label{eq:coupled_sys_2}
\cA(f^h_2,g^h_2) &= \cQ(g^h_2)- \cB(\hat{f}^{h,*}_1,g^h_2)\:,\\
\label{eq:coupled_proj_1}
\hat{f}^{h,*}_1  &= \cP_1(f^h_1)\:,\\
\label{eq:coupled_proj_2}
\hat{f}^{h,*}_2  &= \cP_2(f^h_2)\:,
\end{align}
\end{subequations}
where $\cB(\hat{f}^{h,*}_1,g^h_2):=\cA(f^h_1,g^h_2)$, $\cB(\hat{f}^{h,*}_2,g^h_1):=\cA(f^h_2,g^h_1)$, and the equations in \eqref{eq:coupled_sys_1} and \eqref{eq:coupled_sys_2} are coupled only through the projections in \eqref{eq:coupled_proj_1} and \eqref{eq:coupled_proj_2}.
Suppose that $\hat{f}^{h,*}_1$ and $\hat{f}^{h,*}_2$ are known; then \eqref{eq:coupled_sys_1} and \eqref{eq:coupled_sys_2} are fully decoupled.
In each subdomain ($\Dom_1$ or $\Dom_2$) of the phase space, the sign of $v$ is fixed and  only $E$ is allowed to change sign.
Thus, there is no cyclic dependency in these subdomains, and the decoupled systems \eqref{eq:coupled_sys_1} and \eqref{eq:coupled_sys_2} can be solved independently via explicit sweeping approach in $\Dom_1$ and $\Dom_2$, respectively.

Let $\bff$ be the expansion coefficients of $f^h:=f^h_1+f^h_2$ on an orthogonal basis of $\bbP^1(C)$ for all $C\in\cT$, and let $\hat{\bff}^*$ be the expansion coefficients of $\hat{f}^{h,*} := \hat{f}^{h,*}_1 + \hat{f}^{h,*}_2$ on an orthogonal basis of $\bbP^1(e)$ for all $e\in\cF_0$. 
Then \eqref{eq:coupled_sys} can be expressed a linear system in $\bff$ and $\hat{\bff}^*$ as
\begin{subequations}
\label{eq:linear_sys_0}
\begin{align}
\label{eq:linear_sys_1}
A\bff &= \bq- B\hat{\bff}^*\:,\\
\label{eq:linear_sys_2}
\hat{\bff}^* &= P\bff\:,
\end{align}
\end{subequations}
where $\bq$ is the vector of expansion coefficients of the source term $\cQ(g^{h})$, and the matrices $A$, $B$, and $P$ are defined based on the operators $\cA$, $\cB$, $\cP_1$, and $\cP_2$.
See \cite{Garrett-Hauck-2018} for the detailed definitions.
Applying $A^{-1}$ from the left on both sides of \eqref{eq:linear_sys_1} and plugging the resulting equation into \eqref{eq:linear_sys_2} leads to a much smaller linear system
\begin{equation}\label{eq:linear_sys_reduced}
(I+PA^{-1}B)\hat{\bff}^* = PA^{-1}\bq\:,
\end{equation}
where the operation $A^{-1}$ can be performed efficiently via the sweeping approach.
Then $\hat{\bff}^*$ can be computed by solving \eqref{eq:linear_sys_reduced} with a Krylov solver, such as the generalized minimal residual method (GMRES) \cite{Saad-Schultz-1986}.
After obtaining $\hat{\bff}^*$, the full expansion coefficient $\bff$ is then computed by a final sweeping procedure
\begin{equation}\label{eq:last_sweep}
\bff = A^{-1}(\bq - B\hat{\bff}^*)\:.
\end{equation}
To summarize, \eqref{eq:linear_sys_reduced}--\eqref{eq:last_sweep} defines a mapping from the discretized source $\bq$ to the vector $\bff$ which solves the discretized form of the steady-state equation \eqref{eq:steady_state_boundary}.

\subsection{Solving the Poisson equation}
\label{subsec:Poisson}

We solve the Poisson equation \eqref{eq:final_Poisson_eqn} with a continuous Galerkin method with $\bbQ^1$ elements on the same spatial mesh as given in Section~\ref{subsubsec:phase_space}.
Because the method is standard, we omit the details and refer the reader to, for example, \cite{brenner2007mathematical,ern2013theory} for complete presentation.
For a given particle concentration $\rho$, this method maps the Galerkin discretization of $\rho$ to a discretized potential $\Phi$, and, since $\bbQ^1$ elements are used, the discretized electric field $E$ can be directly calculated from $\Phi$.

\subsection{Solving the drift-diffusion equation}
\label{subsec:drift-diffusion}

In the drift-diffusion limit, numerically solving the scaled semiconductor kinetic equation \eqref{eq:final_semiconductor_eqn} is difficult since the system is stiff.  
As discussed in Section~\ref{subsubsec:low_field_scalings_DD}, the drift-diffusion equation \eqref{eq:drift_diffusion_eqn} serves as a good approximation to \eqref{eq:final_semiconductor_eqn} near the drift-diffusion limit. 
To accelerate the solution procedure of \eqref{eq:final_semiconductor_eqn} near this limit, we apply a synthetic acceleration \cite{Adams-Larsen-2002,Larsen-Morel-2010} and use the solution to \eqref{eq:drift_diffusion_eqn} as a preconditioner when solving \eqref{eq:final_semiconductor_eqn}.  
The detailed discussion of this acceleration technique is given in Section~\ref{subsec:acce_precond}.  
Here we focus on the discretization of \eqref{eq:drift_diffusion_eqn}.  
Discretizing in time with backward Euler gives
\begin{equation}
\label{eq:DD_eqn_BE}
\frac{\xi}{\Delta t} {\varrho^{n+1}}  -  \p_{{x}} {\left( \omega^{-1} \p_{{x}} {\varrho^{n+1}} \right)} +  \beta^2 \p_{{x}} { \left( \omega^{-1} E^{n+1} {\varrho^{n+1}} \right) } = \frac{\xi}{\Delta t}  {\varrho^{n}}\:.
\end{equation}
If $E^{n+1}$ is known a priori at time $t^n$, then \eqref{eq:DD_eqn_BE} takes the steady-state form
\begin{equation}
\label{eq:DD_eqn_BE_simple}
 - \p_{{x}} {\left( \omega^{-1} \p_{{x}} {\varrho} \right)} + \beta^2 \p_{{x}} { \left( \omega^{-1} E {\varrho} \right) } + \frac{\xi}{\Delta t} {\varrho} = Q\:,
\end{equation}
where $E$ is a given electric field and $Q$ is a general source. 

We solve \eqref{eq:DD_eqn_BE_simple} with the direct Discontinuous Galerkin method with interface correction (DDG-IC).
The original DDG scheme \cite{Liu2009jc} is derived based on the weak formulation of \eqref{eq:DD_eqn_BE_simple} with numerical fluxes approximating derivatives of the solution at element boundaries.
The interface correction for DDG was introduced later in \cite{Liu2010eq} to obtain the optimal $(k+1)$-th order of accuracy for polynomial approximations of degree $k$.
It is shown in \cite{{Chen2016gk}} that, with proper choices of numerical flux and limiters, the DDG-IC method satisfies the maximum principle with accuracy up to third order.

Let the spatial domain $[0,L]$ be divided into $N_x$ cells $\{I_j \}_{j=1}^{N_x}$, where $I_j = [x_{j-1},x_j]$ with $x_j = j \dx$, and let $\cV^h = \{\varphi^h\in  L^2 (\bbR) \colon$ $\varphi^h|_{[0,L]^c} = 0 \:,\, \varphi^h|_{I_j} \in \bbP^1(I_j)\:,\,\forall j\ = 1, \dots, N_x\}$ denote the numerical solution space.
For $\varphi^h\in \cV^h$, we define the numerical trace of $\varphi^h$ at cell interface $x_j$ as
$\varphi^{h,\pm}_j := \lim_{\epsilon\to0^+}\varphi^h(x_j\pm\epsilon)$ for $j=0,\dots,N_x$.
The jump and average of $\varphi^h$ at $x_j$ are defined respectively as
\begin{equation}
[ \varphi^{h} ]_{j} = \varphi^{h,+}_j-\varphi^{h,-}_j
\quand
(\overline{\varphi^{h}})_j = \half\big(\varphi^{h,+}_j + \varphi^{h,-}_j\big)\:.
\end{equation}
The DDG-IC scheme then solves \eqref{eq:DD_eqn_BE_simple} by finding the solution $\varrho^h \in \cV^h$ such that, for any test function $\varphi^h \in \cV^h$ and on any cell $I_j$, 
\begin{multline}
\label{eqn:DD_DDGIC}
\int_{I_j} \tau^h\p_x {\varrho^h} \p_x {\varphi^h}\, \dd x 
- {{\varphi^{h,-}_j}}  (  \widehat{\tau^h\p_x{\varrho^h}} )_j  
+ {{\varphi^{h,+}_{j-1}}}  (  \widehat{\tau^h\p_x{\varrho^h}} )_{j-1}
+ \big(\overline{\p_x {\varphi^h}}\big)_{j}  [\tau^h\p_x{\varrho^h}]_{j} 
+ \big(\overline{\p_x {\varphi^h}}\big)_{j-1}  [\tau^h\p_x{\varrho^h}]_{j-1}\\  
-\int_{I_j} \tau^h{\beta^2} E {\varrho^h}  \p_x {\varphi^h}\, \dd x   
+ \varphi^{h,-}_{j}   ( \widehat{\tau^h E{\varrho^h}})_j  
- \varphi^{h,+}_{j-1} ( \widehat{\tau^h E{\varrho^h}})_{j-1}
+\frac{\xi}{\dt}  \int_{I_j} {{\varrho^h}} {\varphi^h}\, \dd x 
 = \int_{I_j} Q  {\varphi^h}\, \dd x\:,
\end{multline}
where $\tau^h$ is the $L^2$ orthogonal projection of $\omega^{-1}$ onto $\cV^h$, and the fourth and fifth terms in \eqref{eqn:DD_DDGIC} are the interface correction terms.
Here the numerical flux $\widehat{\tau^h\p_x{\varrho^h}}$ at $x_j$ is defined as
\begin{equation}
 (\widehat{\tau^h\p_x{\varrho^h}})_{j}   = \frac{2}{\dx} [\tau^h{\varrho^h}]_j + (\overline{\tau^h\p_x{\varrho^h}})_j=
 \frac{2}{\dx}\big(\tau^{h,+}_j \varrho^{h,+}_j - \tau^{h,-}_j \varrho^{h,-}_j\big) + \frac{1}{2}\big(\tau^{h,+}_j (\p_x\varrho)_j^{h,+} + \tau^{h,-}_j (\p_x\varrho)^{h,-}_j\big)   \:,
\end{equation} 
and the Lax-Friedrich flux is used for $\widehat{\tau^hE \varrho^h}$, i.e.,
\begin{equation}
 (\widehat{\tau^h E {\varrho^h}})_{j}  = \half \big( E_j \tau^{h,-}_j \varrho^{h,-}_{j} + E_{j+1}\tau^{h,+}_j \varrho^{h,+}_{j} - \alpha_j[\varrho^h]_{j} \big)\:,\quad \alpha_j:=\max\{|E_j\tau^{h,-}_j|,|E_{j+1}\tau^{h,+}_j|\}\:,
\end{equation} 
where $E_j$ and $E_{j+1}$ are the values of $E$ at the cell centers of $I_j$ and $I_{j+1}$, respectively.
Here the term $\omega^{-1}$ is not involved in the numerical fluxes since the collision frequency $\omega$ is assumed to be known on the entire spatial domain, including the cell boundaries.

\section{Nonlinear solution strategies}
\label{sec:techniques}
In this section, we propose several strategies for solving \eqref{eq:final_semiconductor_model}.
To simplify the discussion, we first introduce a concise operator notation for the fast sweeping method discussed in Section \ref{subsubsec:fast_sweeping}.
Specifically, we write \eqref{eq:backward_euler} as
\begin{equation}
\label{eq:operator_form_implicit}
\cL_{{E}^{n+1}} f^{n+1} = \cS{\rho^{n+1}} + {s}\:,
\end{equation}
where
\begin{equation}
\label{eq:operator_def}
\cL_{E} f := v \p_x f + \beta^2 E \p_v f + \left(\frac{\delta}{\dt}+\frac{\omega}{\epsilon}\right) f\:,\quad
\cS \rho:=\frac{\omega}{\epsilon} \rho M_{\alpha^2}\:,\quand 
{s}:=\frac{\delta}{\dt}f^n\:.
\end{equation}
We refer to $\cS\rho$ and $s$ as the scattering source and the general source, respectively.
If $E^{n+1}=\Wtilde{E}$ and $\rho^{n+1}=\Wtilde{\rho}$, where $\Wtilde{E}$ and $\Wtilde{\rho}$ are known,
then $f^{n+1}$ satisfies a steady-state problem
of the form \eqref{eq:steady_state}:
\begin{equation}
\label{eq:operator_form_steady_state}
\cL_{\Wtilde{E}} f = \cS\Wtilde{\rho} + {s}\:.
\end{equation}
We let 
\begin{equation}
\label{eq:operator_form_discrete}
\hL_{\WEh} \fh =   \hS\Wrhoh + \sh
\end{equation}
denote the discretization of \eqref{eq:operator_form_steady_state} as described in Section~\ref{subsubsec:phase_space}.
Here the operators $\hL$ and $\hS$ are discretized versions of $\cL$ and $\cS$ while $\fh$, $\WEh$, $\Wrhoh$, and $\sh$ denote the discretizations of $f$, $\Wtilde{E}$, $\Wtilde{\rho}$, and $s$, respectively.
The solver discussed in Section~\ref{subsubsec:fast_sweeping} then computes $\fh^{n+1}$ by solving the linear problem
\begin{equation}
\label{eq:operator_sweep}
\fh = \hL_{\WEh}^{-1} (  \hS\Wrhoh + \sh )\:,
\end{equation}
where the operation $\hL_{\WEh}^{-1}$ is performed using the sweeping algorithm from \cite{Garrett-Hauck-2018} that is summarized in Section~\ref{subsubsec:fast_sweeping}.

In the self-consistent setting, $\rho^{n+1}:=\int_\bbR f^{n+1}\dd v$, and $E^{n+1}$ is coupled to $\rho$ via the Poisson equation \eqref{eq:final_Poisson_eqn}.
Thus, instead of the linear problem \eqref{eq:operator_sweep}, we solve
\begin{equation}
\label{eq:solve_f_FP}
\fh = \hL_{\Eh}^{-1} (  \hS\hP\fh + \sh )\:,\quad \Eh = \hF(\hP\fh)\:,
\end{equation}
where $\hP$ denotes integration over the computational velocity domain $[a_v,\,b_v]$ and $\hF$, which maps a given particle concentration to an electric field, denotes the solution procedure of the Poisson equation \eqref{eq:final_Poisson_eqn} using the continuous Galerkin method in Section~\ref{subsec:Poisson}.
This problem is nonlinear since $\Eh$ depends on $\fh$.

The problem \eqref{eq:solve_f_FP} can be solved via nonlinear fixed-point iterative solvers.
However, the cost of solving \eqref{eq:solve_f_FP} could be prohibitive in the multi-dimensional setting due to the high dimensionality of $\fh$.
To reduce the problem dimension, a common trick (see, e.g., \cite{Lewis-Miller-1984,Adams-Larsen-2002}) is to integrate the first equation in \eqref{eq:solve_f_FP} with respect to $v$ and solve the resulting fixed-point problem for $\rhoh$ and $\Eh$:
\begin{equation}
\label{eq:solve_rho_FP}
\rhoh = \hP\hL_{\Eh}^{-1}(\hS\rhoh + \sh)\:,\quad \Eh=\hF(\rhoh)\:.
\end{equation}
The solution to \eqref{eq:solve_rho_FP} gives $\rhoh^{n+1}$ and $\Eh^{n+1}$.
Thus $\fh^{n+1}$ can be computed by a final sweeping procedure by setting $\WEh=\Eh^{n+1}$ and $\Wrhoh=\rhoh^{n+1}$ in \eqref{eq:operator_sweep}.
In the remainder of this section, we consider two different formulations of \eqref{eq:solve_rho_FP}.

\subsection{Type-\Rmnum{1} formulation}
\label{subsec:coupled}

The type-\Rmnum{1} approach formulates \eqref{eq:solve_rho_FP} as a nonlinear fixed-point problem on $\rhoh$, i.e.,
\begin{equation}
 \rhoh = \hG_1(\rhoh) := \hP\hL_{\hF(\rhoh)}^{-1}(\hS \rhoh + \sh) 
 \label{eq:coupled_FP}
\end{equation}
where $\hG_1$ is nonlinear due to the coupling between $\Eh(=\hF(\rhoh))$ and $\rhoh$.
To solve \eqref{eq:coupled_FP}, two iterative solvers are considered: standard Picard iteration (see, e.g.,  \cite[Section~I.8]{hairer1993solving}) and Anderson acceleration \cite{Anderson-1965,Walker-Ni-2011}.

\subsubsection{Type-\Rmnum{1} -- Picard iteration}
\label{subsubsec:coupled_SI}
With an initial guess $\rhoh^{(0)}$, Picard iteration lags both the scattering source and the electric field terms in $\hG_1$ and updates the electron concentration by evaluating $\hG_1$.
Specifically, the Picard iteration update at iteration $k+1$ is given by  
\begin{equation}
\label{eq:coupled_SI}
\rhoh^{(k+1)} = \hG_1(\rhoh^{(k)}) = \hP \hL_{\hF(\rhoh^{(k)})}^{-1} (\hS \rhoh^{(k)} + \sh)\:.
\end{equation}
It is well-known that Picard iteration converges when $\hG_1$ is a contraction mapping, and the rate of convergence depends on the spectral radius of the Jacobian of $\hG_1$.

\subsubsection{Type-\Rmnum{1} -- Anderson acceleration}
\label{subsubsec:coupled_AA}
Anderson acceleration was first proposed in \cite{Anderson-1965} as an acceleration method based on nonlinear Krylov solvers for fixed-point problems.
Here we adopt the variant given in \cite{Walker-Ni-2011} for solving \eqref{eq:coupled_FP}.
At iteration $k+1$, Anderson acceleration first computes the residual
\begin{equation}
\label{eq:AA_1}
\hh_1^{(k)} := \hG_1(\rhoh^{(k)}) - \rhoh^{(k)}\:,
\end{equation}
then solves the least-squares problem
\begin{equation}
\label{eq:AA_2}
\alpha^*:=\argmin_{\alpha\in\bbR^{m_k+1}} \bigg\{ \bigg\|\sum_{i=0}^{m_k} \alpha_i \hh_1^{(k-i)}  \bigg\|_2^2\,\,
\colon\,\, \sum_{i=0}^{m_k} \alpha_i = 1 \bigg\} 
\end{equation}
with $m_k:=\min\{m,k\}$, and finally updates 
\begin{equation}
\label{eq:AA_3}
\rhoh^{(k+1)} = \sum_{i=0}^{m_k} \alpha^*_i \hG_1(\rhoh^{(k-i)})\:.
\end{equation}
Here the truncation parameter $m$ is a nonnegative integer that indicates the maximum number of residuals maintained in memory.
When $m=0$,  Anderson acceleration reduces to standard Picard iteration.
For $m>0$, Anderson acceleration updates $\rhoh$ with a convex combination of the previous $m_k$ iterates that leads to the minimum residual.
It is proved in \cite{Toth-Kelley-2015} that Anderson acceleration converges if Picard iteration converges. 
Since Anderson acceleration utilizes information from the previous iterations, it is expected to converge faster than Picard iteration in practice, but at a cost of additional memory usage.
When the problem is linear, Anderson acceleration has been shown to be equivalent to GMRES under some mild assumptions \cite{Walker-Ni-2011}.

\subsection{Type-\Rmnum{2} formulation}
\label{subsec:decoupled}

The type-\Rmnum{2} fixed-point formulation of \eqref{eq:solve_rho_FP} aims to reduce the nonlinearity of the type-\Rmnum{1} formulation \eqref{eq:coupled_FP} by relaxing the coupling between the electric field $\Eh$ and electron concentration $\rhoh$ in the iterative procedure.
In other words, the type-\Rmnum{2} formulation gives fixed-point problems that can be solved by a nested iterative procedure that consists of a nonlinear outer loop on $\Eh$ and a linear inner loop on $\rhoh$.
The intent is that the inner loop will provide a fast, accurate update of $\rhoh$ to feed into the outer loop, thereby improving the overall efficiency.
To derive the type-\Rmnum{2} formulation, we write \eqref{eq:solve_rho_FP} as
\begin{equation}
(\hI-\hP\hL_{\Eh}^{-1}\hS)\rhoh = \hP\hL_{\Eh}^{-1} \sh\:,\quad \Eh=\hF(\rhoh)\:,
\end{equation}
where $\hI$ is the identity operator. 
Thus the nonlinear fixed-point problem on $\Eh$ takes the form
\begin{equation}
 \label{eq:decoupled_E_FP}
\Eh = \hF(\rhoh) = \hF\big( (\hI-\hP\hL_{\Eh}^{-1}\hS)^{-1} \hP\hL_{\Eh}^{-1} \sh \big)\:,
\end{equation}
where the right-hand side involves solving a linear system. 
To formulate a fixed-point on $\rho$ as the type-\Rmnum{1} problem \eqref{eq:coupled_FP}, the type-\Rmnum{2} formulation takes the following equivalent form of \eqref{eq:decoupled_E_FP}:
\begin{equation}
 \rhoh = \hG_2(\rhoh) := (\hI - \hP\hL_{\hF(\rhoh)}^{-1}\hS )^{-1}\hP\hL_{\hF(\rhoh)}^{-1} \sh \:.
 \label{eq:decoupled_FP}
\end{equation}
Here $\hG_2(\rhoh)$ depends on $\rhoh$ only through the electric field $\hF(\rhoh)$, and evaluating $\hG_2(\rhoh)$ also requires solving the linear system.
Specifically, to compute $\hG_2(\Wrhoh)$ for a given $\Wrhoh$, we solve
\begin{equation}
(\hI - \hP\hL_{\hF(\Wrhoh)}^{-1}\hS ) \hG_2(\Wrhoh) = \hP\hL_{\hF(\Wrhoh)}^{-1} \sh\:.
\label{eq:decoupled_LS}
\end{equation}
Nonlinear iterative solvers are used in the nested procedure to solve \eqref{eq:decoupled_FP} in the outer loop, while linear iterative solvers are considered for solving \eqref{eq:decoupled_LS} in the inner loop.
Both Picard iteration and Anderson acceleration can serve as the nonlinear solver in the outer loop.
However, we only consider Anderson acceleration in this paper, since preliminary numerical results indicate that using Picard iteration here is not competitive in terms of computation time.
For clarity, Anderson acceleration for solving \eqref{eq:decoupled_FP} is stated here.
At iteration $k+1$ in the outer loop, Anderson acceleration first computes the residual
\begin{equation}
\label{eq:AA_21}
\hh_2^{(k)} := \hG_2(\rhoh^{(k)}) - \rhoh^{(k)}\:,
\end{equation}
then solves the least-squares problem
\begin{equation}
\label{eq:AA_22}
\alpha^*:=\argmin_{\alpha\in\bbR^{m_k+1}} \bigg\{ \bigg\|\sum_{i=0}^{m_k} \alpha_i \hh_2^{(k-i)}  \bigg\|_2^2\,\,
\colon\,\, \sum_{i=0}^{m_k} \alpha_i = 1 \bigg\} 
\end{equation}
with $m_k:=\min\{m,k\}$, and finally updates
\begin{equation}
\label{eq:AA_23}
\rhoh^{(k+1)} = \sum_{i=0}^{m_k} \alpha^*_i \hG_2(\rhoh^{(k-i)})\:.
\end{equation}
To evaluate $\hG_2(\rhoh^{(k)})$ in \eqref{eq:AA_21}, we replace $\Wrhoh$ with $\rhoh^{(k)}$ in \eqref{eq:decoupled_LS} and solve the resulting linear system using either Picard iteration or GMRES \cite{Saad-Schultz-1986}.
These iterative solvers form the inner loop of the nested procedure.
In the following subsections, we discuss the details of the application of Picard iteration and GMRES in the inner loop.

\subsubsection{Type-\Rmnum{2} -- Anderson acceleration with Picard iteration}
\label{subsubsec:decoupled_SI}

Let $\rhoh^{(k)}$ denote the $k$-th iterate in the outer loop. 
To evaluate $\hG_2(\rhoh^{(k)})$, we apply Picard iteration on an equivalent fixed-point formulation of \eqref{eq:decoupled_LS}.
Specifically, at iteration $k+1$ in the outer loop, Picard iteration updates
\begin{equation}
\label{eq:decoupled_SI}
\rhoh^{(k,\ell+1)} =  \hP\hL_{\hF(\rhoh^{(k)})}^{-1}(\hS \rhoh^{(k,\ell)} + \sh)\:,
\end{equation}
where $\rhoh^{(k,0)}=\rhoh^{(k)}$.
Let $\rhoh^{(k,*)}:=\lim_{\ell\to\infty} \rhoh^{(k,\ell)}$ denote the limit point of iterates generated by \eqref{eq:decoupled_SI}, then it follows that $\hG_2(\rhoh^{(k)}) = \rhoh^{(k,*)}$.

\subsubsection{Type-\Rmnum{2} -- Anderson acceleration with GMRES}
\label{subsubsec:decoupled_GMRES}

Let $\rhoh^{(k)}$ still denote the $k$-th iterate in the outer loop.
$\hG_2(\rhoh^{(k)})$ can also be evaluated by solving the linear system
\begin{equation}
(\hI- \hP\hL_{\hF(\rhoh^{(k)})}^{-1}\hS) \hG_2(\rhoh^{(k)}) =  \hP\hL_{\hF(\rhoh^{(k)})}^{-1} \sh 
 \label{eq:decoupled_GMRES}
\end{equation}
using GMRES.
In general, the GMRES solver is expected to converge in fewer iterations than the Picard iteration in \eqref{eq:decoupled_SI}.
We will verify this in the numerical results reported in Section~\ref{sec:num_results}.

\subsection{Synthetic acceleration for semiconductor equations}
\label{subsec:acce_precond}

Synthetic acceleration (SA) schemes were first developed in \cite{lebedev1964iterativeKP,kopp1963synthetic} to improve efficiency of iterative solvers for transport equations.
The basic idea of these schemes is to compute a coarse, cheap, or low-order correction term from residuals of the base iterative solver, and apply this correction to the current iterate to accelerate convergence of the base solver.
As noted in \cite{Larsen1991}, many of the synthetic acceleration schemes can be formulated as two-level multigrid algorithms.

In Section~\ref{subsubsec:DSA}, we derive an SA scheme for the semiconductor equation \eqref{eq:final_semiconductor_eqn}, and then apply this scheme to both type-\Rmnum{1} and type-\Rmnum{2} Picard iteration solvers considered in Sections~\ref{subsubsec:coupled_SI} and \ref{subsubsec:decoupled_SI}. 
In Section~\ref{subsubsec:precond}, we follow the same approach as in \cite{Adams-Larsen-2002,derstine1985use,Bruss-Morel-Ragusa-2014} to formulate the synthetic acceleration as a preconditioner.
We then accelerate the type-\Rmnum{1} and type-\Rmnum{2} Krylov solvers in Sections~\ref{subsubsec:coupled_AA} and \ref{subsubsec:decoupled_GMRES} by applying these solvers on the preconditioned problems.

\subsubsection{SA scheme on Picard iteration}
\label{subsubsec:DSA}

The derivation of SA schemes for steady-state linear transport equations with neutral particles can be found in, for example, \cite[Section~II.B.]{Adams-Larsen-2002} and \cite[Section~2-3]{Lewis-Miller-1984}. 
These equations are well approximated in collisional by diffusion equations, which are used to compute cheap corrections to iterates in a solver, resulting in the diffusion synthetic acceleration (DSA) scheme \cite{Alcouffe-1976,Alcouffe-1977,Adams-Larsen-2002,Lewis-Miller-1984}. 

For the semiconductor equations, it is known \cite{Markovich-Ringhofer-Schmeiser-1990,Poupaud-1991,Masmoudi-Tayeb-2007} that drift-diffusion equations serve as proper low-order approximations to the semiconductor equations in the drift-diffusion limit, as discussed in Section~\ref{subsubsec:low_field_scalings_DD}. 
In this section, we derive an SA scheme for semiconductor equations with the correction term computed by solving a drift-diffusion equation.
The derivation is mostly a straightforward extension of the derivation for the DSA scheme in \cite{Adams-Larsen-2002}, and we include it here for completeness.

To derive the SA scheme, we first rewrite the semi-discrete semiconductor equation \eqref{eq:operator_form_implicit} as a steady-state equation 
\begin{equation}
\label{eq:DSA_1}
\cL_{\cF(\cP f)}  f = \cS\cP f + s \:.
\end{equation}
Here $\cP$ denotes integration over the velocity domain and $\cF$ denotes the mapping from the particle concentration $\rho=\cP f$ to the electric field $E=\cF(\cP f)$ via the Poisson equation \eqref{eq:final_Poisson_eqn}.
Applying $\cL_{\cF(\cP f)}^{-1}$ on both sides of \eqref{eq:DSA_1} and solving the resulting equation with Picard iteration leads to
\begin{equation}
\label{eq:DSA_2}
 f^{(k+\shalf)} = \cL_{\cF(\cP f^{(k)})}^{-1} (\cS\cP f^{(k)} + s )\:,
\end{equation}
where index of the update is now $k+\half$ instead of $k+1$. 
Integrating \eqref{eq:DSA_2} with respect to $v$ gives 
\begin{equation}
\label{eq:DSA_2_1}
\rho^{(k+\shalf)} = \cP\cL_{\cF(\rho^{(k)})}^{-1} (\cS\rho^{(k)} + s )\:,
\end{equation}
which is a continuous version of the type-\Rmnum{1} Picard iteration \eqref{eq:coupled_SI}.
To derive a correction for \eqref{eq:DSA_2_1}, we write \eqref{eq:DSA_2} as
\begin{equation}
\label{eq:DSA_3}
\cL_{\cF(\cP f^{(k)})} f^{(k+\shalf)} = \cS\cP f^{(k)} + s \:,
\end{equation}
and subtract \eqref{eq:DSA_3} from \eqref{eq:DSA_1}.
By adding and subtracting terms in the resulting equation, we have
\begin{equation}
\label{eq:DSA_5}
(\cL_{\cF(\cP f)}   - \cL_{\cF(\cP f^{(k)})} ) f + \cL_{\cF(\cP f^{(k)})}  \psi - \cS\cP \psi = \cS\cP (f^{(k+\shalf)} - f^{(k)}) \:,
\end{equation}
where $\psi:=f-f^{(k+\shalf)}$ denotes the error.
If $\psi$ can be computed by solving \eqref{eq:DSA_5}, then it can be used as a correction to Picard iterates by taking $f^{(k+1)} = f^{(k+\shalf)} + \psi$. 
However, solving \eqref{eq:DSA_5} is equivalent to finding $f$.

The SA scheme considered here computes corrections to Picard iterates by solving a reduced-order equation that approximates \eqref{eq:DSA_5}.
To obtain the reduced-order equation, we apply $\cP$ to both sides of \eqref{eq:DSA_5}, which leads to
\begin{equation}
\label{eq:DSA_6}
\cP \cL_{\cF(\rho^{(k)})}  \psi - \cP \cS \phi = \cP\cS (\rho^{(k+\shalf)} - \rho^{(k)}) \:,
\end{equation}
where $\phi:=\cP \psi$ is the integral of $\psi$ over the velocity domain.
Here the first term in \eqref{eq:DSA_5} vanishes since, for any electron distribution $g$ and any electric fields $E_1$ and $E_2$, it follows from \eqref{eq:operator_def} that $\cP(\cL_{E_1}- \cL_{E_2})g = \beta^2 (E_1-E_2)\int_\bbR \p_v g \dd v = 0$, provided that $g$ goes to zero as $v\to\pm\infty$.
Motivated by the drift-diffusion limit, we approximate the operators on the left-hand side of \eqref{eq:DSA_6} by a drift-diffusion operator $\cD_{E}$ defined as
\begin{equation}
\label{eq:DSA_DD_operator}
\cD_{E} \phi = - \epsilon\p_x(\omega^{-1} \p_x\phi) + \epsilon\beta^2\p_x(\omega^{-1} E \phi) + \left(\frac{\delta}{\dt}\right) \phi\:.
\end{equation}
Thus, \eqref{eq:DSA_6} is approximated by
\begin{equation}
\label{eq:DSA_DD}
\cD_{\cF(\rho^{(k)})}\Wtilde{\phi} = \cP\cS (\rho^{(k+\shalf)} - \rho^{(k)}) \:.
\end{equation}
We correct the Picard iterate in \eqref{eq:DSA_2_1} with $\Wtilde{\phi}$, a solution to \eqref{eq:DSA_DD}, by taking $\rho^{(k+1)} = \rho^{(k+\shalf)} + \Wtilde{\phi}$, which results in the SA scheme 
\begin{subequations}
\label{eq:SI_DSA}
\begin{align}
\label{eq:SI_DSA_1}
{\rho}^{(k+\shalf)} &= \cP \cL_{\cF(\rho^{(k)})}^{-1} (\cS \rho^{(k)} + s)\:,\\
\label{eq:SI_DSA_2}
\rho^{(k+1)} &= {\rho}^{(k+\shalf)} + \cD_{\cF(\rho^{(k)})}^{-1}  \cP \cS ({\rho}^{(k+\shalf)}-\rho^{(k)})   \:,
\end{align}
\end{subequations}
where \eqref{eq:SI_DSA_2} follows from the definition of $\Wtilde{\phi}$ in \eqref{eq:DSA_DD}.
The SA scheme based on the type-\Rmnum{1} Picard iteration \eqref{eq:coupled_SI} is then given by a discretized version of \eqref{eq:SI_DSA}
\begin{subequations}
\label{eq:coupled_SI_DSA}
\begin{align}
\label{eq:coupled_SI_DSA_1}
{\rhoh}^{(k+\shalf)} &= \hP \hL_{\hF(\rhoh^{(k)})}^{-1} (\hS \rhoh^{(k)} + \sh)\:,\\
\label{eq:coupled_SI_DSA_2}
\rhoh^{(k+1)} &= {\rhoh}^{(k+\shalf)} + \hD_{\hF(\rhoh^{(k)})}^{-1}  \hP \hS ({\rhoh}^{(k+\shalf)}-\rhoh^{(k)})   \:,
\end{align}
\end{subequations}
where $\hD_{\hF(\rhoh^{(k)})}^{-1}$ denotes the solution procedure of the drift-diffusion equation \eqref{eq:DSA_DD} using the DDG-IC solver presented in Section~\ref{subsec:drift-diffusion}.

An analogous SA scheme based on the type-\Rmnum{2} Picard iteration \eqref{eq:decoupled_SI} can be derived by repeating the analysis above with minor modification in the indices in \eqref{eq:DSA_2}. 
The resulting SA takes the form
\begin{subequations}
\label{eq:decoupled_SI_DSA}
\begin{align}
\rhoh^{(k,\ell+\shalf)} &=  \hP\hL_{\hF(\rhoh^{(k)})}^{-1}(\hS \rhoh^{(k,\ell)} + \sh)\:,\\
\rhoh^{(k,\ell+1)} &= \rhoh^{(k,\ell+\shalf)} + \hD_{\hF(\rhoh^{(k)})}^{-1}  \hP \hS (\rhoh^{(k,\ell+\shalf)}-\rhoh^{(k,\ell)})   \:,
\end{align}
\end{subequations}
where $\hD_{\hF(\rhoh^{(k)})}^{-1}$ is still performed using the DDG-IC solver in Section~\ref{subsec:drift-diffusion}.

\subsubsection{Preconditioner form of the SA scheme}
\label{subsubsec:precond}
The SA scheme in Section~\ref{subsubsec:DSA} is derived specifically for Picard iteration.
It is well-known \cite{Bruss-Morel-Ragusa-2014,derstine1985use,Adams-Larsen-2002} that many SA schemes can be formulated as preconditioners.
Following this approach, we derive the SA scheme in the preconditioner forms for the type-\Rmnum{1} Anderson acceleration and the type-\Rmnum{2} GMRES solver considered in Sections~\ref{subsubsec:coupled_AA} and \ref{subsubsec:decoupled_GMRES}, respectively.

We first consider the type-\Rmnum{1} Anderson acceleration in Sections~\ref{subsubsec:coupled_AA}.
In this case, the SA scheme applies Anderson acceleration \eqref{eq:AA_1}--\eqref{eq:AA_3} to a preconditioned version of fixed-point problem \eqref{eq:coupled_FP}.
This preconditioned problem is derived from the SA scheme based on type-\Rmnum{1} Picard iteration \eqref{eq:coupled_SI_DSA} by first rewriting the correction process \eqref{eq:coupled_SI_DSA_2} in the residual form as
\begin{equation}
\label{eq:DSA_residual}
\rhoh^{(k+1)} - \rhoh^{(k)} = (\hI + \hD_{\hF(\rhoh^{(k)})}^{-1}  \hP \hS) (\rhoh^{(k+\shalf)} - \rhoh^{(k)}) \:.
\end{equation}
Plugging \eqref{eq:coupled_SI_DSA_1} into \eqref{eq:DSA_residual} then gives
\begin{equation}
\label{eq:DSA_residual_2}
\rhoh^{(k+1)} - \rhoh^{(k)} = -(\hI + \hD_{\hF(\rhoh^{(k)})}^{-1}  \hP \hS) \big( (\hI - \hP \hL_{\hF(\rhoh^{(k)})}^{-1}\hS) \rhoh^{(k)} - \hP \hL_{\hF(\rhoh^{(k)})}^{-1} \sh \big) \:,
\end{equation}
which is equivalent to a standard Picard iteration update on the (preconditioned) fixed-point problem
\begin{equation}
\label{eq:precond_FP}
\rhoh =  \Wtilde{\hG}_1(\rhoh) :=
\rhoh - ( \hI + \hD_{\hF(\rhoh)}^{-1}  \hP \hS) \big( (\hI-\hP\hL_{\hF(\rhoh)}^{-1}\hS)\rhoh - \hP\hL_{\hF(\rhoh)}^{-1} \sh \big)\:.
\end{equation}
Here \eqref{eq:precond_FP} is a preconditioned version of \eqref{eq:coupled_FP} with preconditioner $ ( \hI + \hD_{\hF(\rhoh)}^{-1}  \hP \hS)$. 
We obtain the SA scheme based on type-\Rmnum{1} Anderson acceleration by replacing each $\hG_1$ in \eqref{eq:AA_1}--\eqref{eq:AA_3} with $\Wtilde{\hG}_1$.

For the type-\Rmnum{2} GMRES solver in Section~\ref{subsubsec:decoupled_GMRES}, we follow a similar approach and use \eqref{eq:decoupled_SI_DSA} to derive a preconditioned version of the linear system \eqref{eq:decoupled_GMRES}:
\begin{equation}
\label{eq:precond_LS}
( \hI + \hD_{\hF(\rhoh^{(k)})}^{-1}  \hP \hS) (\hI - \hP\hL_{\hF(\rhoh^{(k)})}^{-1}\hS) {\hG}_2(\rhoh^{(k)}) = ( \hI + \hD_{\hF(\rhoh^{(k)})}^{-1}  \hP \hS) \hP\hL_{\hF(\rhoh^{(k)})}^{-1} \sh\:.
\end{equation}
Therefore, the SA scheme based on type-\Rmnum{2} GMRES solver computes $\hG_2(\rhoh^{(k)})$ by solving the preconditioned system \eqref{eq:precond_LS}.

\section{Numerical results}
\label{sec:num_results}

The iterative solvers from the previous section are tested and compared on the one-dimensional silicon $n^+$--$n$--$n^+$ diode problem \cite{Carrillo2000iza,Cercignani2000domain,Cheng2009hm,Cercignani2001iv,Hu-2014} with different collision frequencies.
In Section~\ref{subsec:diode_setup}, we describe the silicon diode problem and state the implementation details.
In Section~\ref{subsec:single_scale}, we consider ``single-scale" problems where the collision frequency is assumed to be constant throughout the spatial domain. 
Results from these single-scale tests illustrate characteristics of the different solvers.
In Section~\ref{subsec:multiscale}, we consider more realistic ``multiscale" problems with collision frequencies varying in the spatial domain.
We first model the diode using the collision frequency specified in \cite{Cercignani2000domain} which depends on the doping profile.
We then consider a more challenging problem with collision frequency that changes more drastically over the spatial domain.

\subsection{Silicon diode problem setup and implementation details}
\label{subsec:diode_setup}

In the silicon (Si) $n^+$--$n$--$n^+$ diode problem, we simulate the electron movement in an one-dimension Si diode of length $L=0.6 \si{\micro m}$ with a bias voltage $V_{\textup{bias}} = 1 \si{V}$.
Here the electron charge, the effective electron mass, and the electric permittivity of the Si material are respectively given by $q_e = 1.602 \times 10^{-19} \si{C}$, $m = 2.368 \times 10^{-31} \si{kg}$, $\varepsilon_p = 1.034 \times 10^{-10} \si{\sfrac{F}{m}}$.
The Boltzmann constant $k_{\textup{B}} = 1.38 \times 10^{-23} \si{\sfrac{J}{K}}$, and the lattice temperature $T=300\si{K}$.
As in \cite{Cercignani2000domain}, for $x \in [0, 0.6] (\si{\micro m})$, the doping profile is
\begin{equation}\label{eq:doping_profile}
D(x) = 
\begin{cases}
2 \times 10^{21}  \si{m^{-3}}, &  x \in [0.1, 0.5] (\si{\micro m}) \\
5 \times 10^{23}  \si{m^{-3}}, &  \textup{otherwise}
\end{cases}\:.
\end{equation}

With these data, we apply the following scaling: $x_0 = 10^{-6} \si{m}$, $[\Phi] = 1\si{V}$, and $D_0 = 10^{21} \si{m^{-3}}$.
We set $\zeta=1$ in \eqref{eq:final_Poisson_eqn} (low-field scaling).
The velocities in \eqref{eq:velocities} take values
\begin{equation}
\Theta^{\sfrac{1}{2}} = 1.322 \times 10^{5} \si{\sfrac{m}{s}}\:,\quad 
B_0 = 8.225 \times 10^{5} \si{\sfrac{m}{s}}\:,\quand 
C_0 = 1.024 \times 10^{6} \si{\sfrac{m}{s}}\:.
\end{equation}
We set the reference velocity to be $v_0 = \max\{\Theta^{\sfrac{1}{2}}, B_0, C_0\} = C_0$.
Thus, the ratios in \eqref{eq:final_semiconductor_model} are $\alpha = 0.129$, $\beta = 0.803$, and $\gamma = 1$.
The nondimensional doping profile on $x \in [0, 0.6]$ is
\begin{equation}\label{eq:D_nondim}
D(x) = 
\begin{cases}
2, &  x \in [0.1, 0.5] \\
500, &  \textup{otherwise}
\end{cases}\:.
\end{equation}
In the numerical simulations, we impose smooth transitions into the doping profile as in \cite{Hu-2014}. 
These smooth transitions are constructed by cubic splines, and the transition regions are of width 0.04, centered at 0.1 and 0.5.

The initial condition and the incoming boundary data for the kinetic equation \eqref{eq:final_semiconductor_eqn} are respectively given by
$f(t_0,x,v) = D(x) M_{\alpha^2}(v)$ and 
\begin{equation}
f(t,0,v) = D(0) M_{\alpha^2}(v)\:,\,\,\forall v>0 \:,\quad f(t,L,v) = D(L) M_{\alpha^2}(v)\:,\,\,\forall v<0\:.
\end{equation}
Without loss of generality, we let $\Phi_0=\Phi(t,0)=0\si{V}$ in the Poisson equation \eqref{eq:final_Poisson_eqn}.
The boundary data for \eqref{eq:final_Poisson_eqn} then become
$\Phi(t,0) = 0 \si{V}$ and $\Phi(t,L) = V_{\textup{bias}}$.
Since $[\Phi] = 1\si{V}$, the scaled boundary data are $\Phi(t,0) = 0$ and $\Phi(t,L) = 1$.

The computation is performed on a truncated domain $\Dom:=[0,0.6]\times[-2,2]$.
The velocity space is truncated from $\bbR$ to $[-2,2]$, i.e., $v_{\max}=2$, since the value of Maxwellian $M_{\alpha^2}(v)$ is smaller than the machine precision when $|v|>2$.
We discretize $\Dom$ into $200\times 50$ uniform rectangular elements of size $\dx\times\dv$, and solve the semiconductor model \eqref{eq:final_semiconductor_model} from initial time $t_0=0$ to final time $t_{\textup{f}}=0.5$, at which point the system is essentially in steady state.
The kinetic equation \eqref{eq:final_semiconductor_eqn} is solved using the fast sweeping algorithm detailed in Section~\ref{subsec:kinetic}, the Poisson equation \eqref{eq:final_Poisson_eqn} is solved via the continuous Galerkin method discussed in Section~\ref{subsec:Poisson}, and the drift-diffusion equation \eqref{eq:DSA_DD}, which is involved in the synthetic acceleration procedure, is solved by the direct discontinuous Galerkin method in Section~\ref{subsec:drift-diffusion}.
The Poisson equation and the drift-diffusion equation are both solved on a uniform mesh with $200$ elements on $[0,0.6]$.
We choose the parameter $m=3$ in Anderson acceleration \eqref{eq:AA_1}--\eqref{eq:AA_3}.
The relative tolerances for the type-\Rmnum{1} and type-\Rmnum{2} iterative solvers are set to $10^{-8}$, while the relative tolerance in the GMRES solver for solving \eqref{eq:linear_sys_reduced} in the fast sweeping algorithm is set to $10^{-10}$.
The lower tolerance on the fast sweeping GMRES solver is due to the fact that it is used to evaluate $\hL^{-1}_\Eh$, which is a fundamental building block in both type-\Rmnum{1} and type-\Rmnum{2} solvers.
The iterative solvers are terminated once the relative residual is below the set tolerance, or when the solvers reach the maximum allowed number of iterations, which is set to $10,000$ for all solvers.  
This number is set to be large solely for studying the behavior of different iterative solvers. 
For practical applications, the maximum allowed number of iterations should be set much lower.

In the following sections, we consider the diode problem described in this subsection with various collision frequencies and compare the performance of the eight iterative solvers introduced in Section~\ref{sec:techniques}.
We also make a formal efficiency comparison of the proposed implicit scheme to standard explicit schemes and standard implicit-explicit (IMEX) asymptotic preserving (AP) schemes.
In all tests, the implicit time step is chosen to be $\dt=\dx$.
We note that the CFL condition for standard explicit schemes takes the form $\dt\leq \min\{C_1 \delta {\dx}/{{v}_{\max}}, C_2 {\delta\epsilon}/{{\omega}_{\max}}\}$ and the CFL condition for standard IMEX-AP schemes takes the form $\dt\leq \max\{C_3 \delta {\dx}/{{v}_{\max}}, C_4 \delta{{\omega}_{\min}}{\dx^2}/\epsilon\}$ \cite{Jin-2012}, where ${v}_{\max}$, ${v}_{\min}$, ${\omega}_{\max}$ and ${\omega}_{\min}$ are the maximum and minimum values of the velocity and collision frequency over the spatial domain, respectively.
For simplicity, we assume that the $O(1)$ constants $C_1$, $C_2$, $C_3$, and $C_4$ are all equal to one.
Here we do not consider the time step restrictions associated to $\dv$, since $\dx$ is much smaller than $\dv$.
Also, the condition $\dt\leq {\delta\epsilon}/{{\omega}_{\max}}$ for standard explicit schemes is never active in problems tested in this section due to the small value of $\dx$.
In the formal comparison, we assume that each explicit or IMEX step roughly requires the same computation time as each sweeping iteration in an implicit time steps.
Thus for each iterative solver, we compute the ratio 
\begin{equation}\label{eq:chi_def}
\bigchi_{\text{EX/IMEX}} := \frac{\text{equivalent number of explicit/IMEX steps per implicit time step}}
{\text{total number of sweeping iterations per implicit time step}}
\end{equation}
as an efficiency indicator of the proposed scheme. 
We also note that parallelization of the explicit or IMEX updates is often possible, while the sweeping procedure in the implicit scheme requires serial implementation. 
We do not take this fact into account in the comparison.

\subsection{Single-scale test}
\label{subsec:single_scale}

In this section, we test the iterative solvers on problems with constant collision frequency on the spatial domain.
We first consider the low collision case, where the electron mobility is approximated (see \cite{Cercignani2000domain}) by 
\begin{equation}
\mu_{\textup{Si}}(x) = 0.0088+\frac{0.1793\times10^{23}}{1.4320\times10^{23}+D(x)}
\end{equation}
with $D(x)=2\times10^{21}\si{m^{-3}}$, which determines the collision frequency $\omega(x)= \frac{q_e}{m \mu_{\textup{Si}}(x)}$.
We scale the collision frequency by $\omega_0=5.114\times10^{12}\si{s^{-1}}$, and the nondimensional collision frequency is $\omega(x)=1$.
The Knudsen number is $\epsilon = \frac{v_0 }{x_0 \omega_0} = 0.200$ and we choose $\delta=\epsilon=0.200$.

Table~\ref{table:single_scale_iter_time_1} reports the iteration counts and computation time for each iterative solver.
Here the column ``FP" gives the iteration counts for solving the fixed-point problems \eqref{eq:coupled_FP} and \eqref{eq:decoupled_FP}, respectively. 
For type-\Rmnum{2} methods, the column ``LS" reports the iteration counts for solving the linear system \eqref{eq:decoupled_LS} when evaluating $\hG_2$ in \eqref{eq:decoupled_FP}.
The column ``SW" gives the iteration counts for the GMRES solver for solving \eqref{eq:linear_sys_reduced} in the fast sweeping algorithm when computing $\hL^{-1}_\Eh$.
As for the solvers, ``PI" and ``AA" stand for Picard iteration and Anderson acceleration, respectively.

From the results reported in Table~\ref{table:single_scale_iter_time_1}, we first observe that the accelerated solvers are slower than the unaccelerated ones.
This result is to be expected, since the system is away from the drift-diffusion limit due to the relatively small collision frequency.
We also observe that the type-\Rmnum{1} solvers are faster than the type-\Rmnum{2} solvers.
We conclude that for the type-\Rmnum{2} solvers, the additional computation cost of solving the linear system \eqref{eq:decoupled_LS} when evaluating $\hG_2$ outweighs the benefit of the more accurate updates for the fixed-point problem
As we expected, the Krylov-type solvers (Anderson acceleration and GMRES) converge in fewer iterations than Picard iteration does. 
For type-\Rmnum{1} solvers, this results in less computation time, while for type-\Rmnum{2} solvers, the higher computation time per iteration of GMRES makes it slower than Picard iteration.

For this problem, the explicit and IMEX-AP CFL conditions both take the form $\dt\leq \delta \dx/ v_{\max}$, which results in time steps that are 10x smaller than the implicit time step $\dt=\dx$ used in the test.
For each tested iterative solver, the value of the efficient indicator $\bigchi$ defined in \eqref{eq:chi_def} is reported in Table~\ref{table:single_scale_iter_eff_1}.
Here Iter$_\text{SW}$ denotes the total number of sweeping iterations required in one implicit step.
From Table~\ref{table:single_scale_iter_eff_1}, the proposed scheme with type-\Rmnum{1} solvers results in $\bigchi>1$ when comparing to both the explicit and IMEX-AP schemes.
When using type-\Rmnum{2} solvers, $\bigchi<1$ as the higher iteration counts outweigh the benefit of larger implicit time steps.

\begin{table}[h]
\centering
\begin{tabular}{|| c || c | r | r | r || c || c | r | r | r | r ||}
\toprule
\multirow{2}{*}{Solver} & \multirow{2}{*}{SA} &
\multicolumn{2}{c|}{Iteration} & {Total} &\multirow{2}{*}{Solver} & \multirow{2}{*}{SA} &
\multicolumn{3}{c|}{Iteration} & {Total}\\
\cline{3-4} \cline{8-10}
 & & {FP} & {SW} & {Time} &  & & {FP} & {LS} & {SW} & {Time} \\
\hline
 {Type-\Rmnum{1}}  	& N & 4.6 & 1.3 &  8.8 & {Type-\Rmnum{2}}  	& N & 3.2 & 3.0 & 1.3 & 21.4 \\
 \cline{2-5}  \cline{7-11}
 {PI}  				& Y & 4.5 & 1.3 &  9.7 & {AA/PI} 				& Y & 3.2 & 2.9 & 1.3 & 22.5 \\
\hline
 {Type-\Rmnum{1}}   	& N & 3.0 & 1.4 &  7.9& {Type-\Rmnum{2}}  	& N & 3.3 & 1.4 & 3.2 & 43.7 \\
 \cline{2-5}  \cline{7-11}
 {AA} 				& Y & 3.3 & 1.3 &  9.2 & {AA/GMRES}  			& Y & 7.3 & 2.1 & 3.3 & 97.8 \\
\bottomrule
\end{tabular}
\caption{Single-scale problem with $\omega(x)=1$, $\delta=\epsilon=0.200$, and $\dt=\dx$. Iteration counts and total computation time (sec) for the compared solvers and their accelerated variants. }
\label{table:single_scale_iter_time_1}
\end{table}

\begin{table}[h]
	\centering
	\begin{tabular}{|| c || c | c | c | c || c || c | c | c | c ||}
		\toprule
		\multicolumn{10}{||c||}{$1$ implicit step $\sim$ $10$ IMEX-AP steps $\sim$ $10$ explicit steps}\\
		\hline
		Solver & SA & {Iter$_{\text{SW}}$} & {$\bigchi_{\text{IMEX}}$} & {$\bigchi_{\text{EX}}$} & Solver & SA & {Iter$_{\text{SW}}$} & {$\bigchi_{\text{IMEX}}$} & {$\bigchi_{\text{EX}}$}\\
		\hline
		{Type-\Rmnum{1}}  	& N & 6.0 & 1.67 & 1.67 & {Type-\Rmnum{2}}  	& N & 12.5 & 0.80 &  0.80 \\
		\cline{2-5}  \cline{7-10}
		{PI}  				& Y & 5.9 & 1.69 & 1.69 & {AA/PI} 			& Y & 12.1 & 0.83 &  0.83 \\
		\hline
		{Type-\Rmnum{1}}   	& N & 4.2 & 2.38 & 2.38 & {Type-\Rmnum{2}}  	& N & 14.8 & 0.68 &  0.68 \\
		\cline{2-5}  \cline{7-10}
		{AA} 				& Y & 4.3 & 2.33 & 2.33 & {AA/GMRES}  		& Y & 50.6 & 0.20 &  0.20 \\
		\bottomrule
	\end{tabular}
	\caption{Single-scale problem with $\omega(x)=1$, $\delta=\epsilon=0.200$, and $\dt=\dx$. The values of the efficiency indicator $\bigchi$ \eqref{eq:chi_def} and the total number of sweeping iterations in an implicit step (Iter$_{\text{SW}}$) are reported. This serves as a formal efficiency comparison between the proposed implicit scheme and the standard explicit and IMEX-AP schemes.}
	\label{table:single_scale_iter_eff_1}
\end{table}

We next test the iterative solvers on problems with large collision frequency that is 100x of the one in the previous problem, and we scale the collision frequency by $\omega_0=5.114\times10^{14}\si{s^{-1}}$ so that the nondimensional collision frequency is still $\omega(x)=1$.
The Knudsen number is then $\epsilon = \frac{v_0 }{x_0 \omega_0} = 0.002$ and again we choose $\delta=\epsilon=0.002$.
We still choose the time step to be $\dt=\dx$.
For this problem, the system is close to the drift-diffusion limit due to the large collision frequency.
Thus we expect that the drift-diffusion based synthetic acceleration would provide sufficient accurate corrections and result in faster convergence for the iterative solvers.
The iteration counts and computation time for this problem are reported in Table~\ref{table:single_scale_iter_time_2}.
From these results, we observe that SA schemes indeed require fewer number of iteration and speed up the base iterative solvers from 1.5x to 8.7x. 
We also note that the computation time per iteration remains roughly the same, which implies that the time spent on computing the drift-diffusion correction term is essentially negligible.

For this problem, the standard explicit time step satisfies $\dt\leq \delta {\dx}/{v_{\max}}$, which is 1000x smaller than the implicit time step $\dt=\dx$.
The standard IMEX-AP time step satisfies $\dt\leq\delta{\omega_{\min}}{\dx^2}/\epsilon$, which is 333x smaller than the implicit time step.
From Table~\ref{table:single_scale_iter_eff_2}, in most cases, the proposed scheme results in $\bigchi>1$ when comparing to both the explicit and IMEX-AP schemes. 
The exceptions are the unaccelerated type-\Rmnum{1} PI solver and the type-\Rmnum{2} AA/PI solvers.

\begin{table}[h]
\centering
\begin{tabular}{|| c || c | r | r | r || c || c | r | r | r | r ||}
\toprule
\multirow{2}{*}{Solver} & \multirow{2}{*}{SA} &
\multicolumn{2}{c|}{Iteration} & \multicolumn{1}{c||}{Total} &\multirow{2}{*}{Solver} & \multirow{2}{*}{SA} &
\multicolumn{3}{c|}{Iteration} & \multicolumn{1}{c||}{Total}\\
\cline{3-4} \cline{8-10}
 & & \multicolumn{1}{c|}{FP} & {SW} & \multicolumn{1}{c||}{Time} &  & & {FP} & \multicolumn{1}{c|}{LS} & {SW} & \multicolumn{1}{c||}{Time} \\
\hline
 {Type-\Rmnum{1}}  	& N &1399.3 & 1.0 & 2011.6 & {Type-\Rmnum{2}}  	& N & 2.5 & 440.1 & 1.0 & 2165.2 \\
 \cline{2-5}  \cline{7-11}
 {PI}  				& Y & 131.9 & 1.0 & 232.9 & {AA/PI} 				& Y & 6.1 & 84.4 & 1.0 &  918.0 \\
\hline
 {Type-\Rmnum{1}}   	& N & 49.7 & 1.0 &  99.9 & {Type-\Rmnum{2}}  	& N & 6.1 & 22.7 & 2.1 &  392.2\\
 \cline{2-5}  \cline{7-11}
 {AA} 				& Y & 29.0 & 1.1 &  67.8 & {AA/GMRES}  			& Y & 7.6 & 9.2 & 2.1 & 218.3 \\
\bottomrule
\end{tabular}
\caption{Single-scale problem with $\omega(x)=1$, $\delta=\epsilon=0.002$, and $\dt=\dx$. Iteration counts and total computation time (sec) for the compared solvers and their accelerated variants.}
\label{table:single_scale_iter_time_2}
\end{table}

\begin{table}[h]
	\centering
	\begin{tabular}{|| c || c | r | r | r || c || c | r | c | c ||}
		\toprule
		\multicolumn{10}{||c||}{$1$ implicit step $\sim$ $333$ IMEX-AP steps $\sim$ $1000$ explicit steps}\\
		\hline
		Solver & SA & {Iter$_{\text{SW}}$} & {$\bigchi_{\text{IMEX}}$} & {$\bigchi_{\text{EX}}$} & Solver & SA & {Iter$_{\text{SW}}$} & {$\bigchi_{\text{IMEX}}$} & {$\bigchi_{\text{EX}}$}\\
		\hline
		{Type-\Rmnum{1}}  	& N & 1399.3 & 0.24 & 0.71 & {Type-\Rmnum{2}}  & N & 1100.3 & 0.30 &  0.91\\
		\cline{2-5}  \cline{7-10}
		{PI}  				& Y & 131.9 & 2.53 & 7.58 & {AA/PI} 			& Y & 514.8 & 0.65 &  1.94 \\
		\hline
		{Type-\Rmnum{1}}   	& N & 49.7 & 6.71 & 20.12 & {Type-\Rmnum{2}}  	& N & 290.8 & 1.15 &  3.44 \\
		\cline{2-5}  \cline{7-10}
		{AA} 				& Y & 32.0 & 10.42& 31.25 & {AA/GMRES}  		& Y & 146.8 & 2.27 &  6.81 \\
		\bottomrule
	\end{tabular}
	\caption{Single-scale problem with $\omega(x)=1$, $\delta=\epsilon=0.002$, and $\dt=\dx$. The values of the efficiency indicator $\bigchi$ \eqref{eq:chi_def} and the total number of sweeping iterations in an implicit step (Iter$_{\text{SW}}$) are reported. This serves as a formal efficiency comparison between the proposed implicit scheme and the standard explicit and IMEX-AP schemes.}
	\label{table:single_scale_iter_eff_2}
\end{table}
 
\subsection{Multiscale test}
\label{subsec:multiscale}

In this section, we test the solvers on multiscale problems with collision frequencies varying in the spatial domain.
The first multiscale problem is the ``standard" silicon diode problem from \cite{Cercignani2000domain,Hu-2014}.
Here the collision frequency is determined by the approximated electron mobility based on the doping profile.
Specifically, with the approximate formula $\mu_{\textup{Si}}(x) = 0.0088+\frac{0.1793\times10^{23}}{1.4320\times10^{23}+D(x)}$ at $T=300\si{K}$ and the doping profile $D(x)$ in \eqref{eq:doping_profile}, the electron mobility is given by
\begin{equation}
\mu_{\textup{Si}}(x) = 
\begin{cases}
0.1323   \si{\frac{m^2}{V s}}, &  x \in [0.1, 0.5] (\si{\micro m}) \\
0.0367   \si{\frac{m^2}{V s}}, &  \textup{otherwise}
\end{cases}\:.
\end{equation}
The nondimensional collision frequency on $x \in [0, 0.6]$ is then
\begin{equation}\label{eq:nu_standard_scaled}
\omega(x) = 
\begin{cases}
0.277, &  x \in [0.1, 0.5] \\
1, &  \textup{otherwise}
\end{cases}
\end{equation}
with the scaling $\omega_0 = 1.843 \times 10^{13}  \si{s^{-1}}$.
Here the Knudsen number $\epsilon = \frac{v_0 }{x_0 \omega_0} = 0.056$, and we choose  $\delta=\epsilon=0.056$.
As mentioned in Section~\ref{subsec:diode_setup}, the doping profile used in the numerical tests includes the smooth transitions as in \cite{Hu-2014}.
The collision frequency at these transition regions are computed from the smooth doping profile using the approximate formula given above.

\begin{figure}[h]
   \captionsetup[subfigure]{justification=centering}
\subfloat[Doping profile]
{\includegraphics[width=0.318\linewidth]{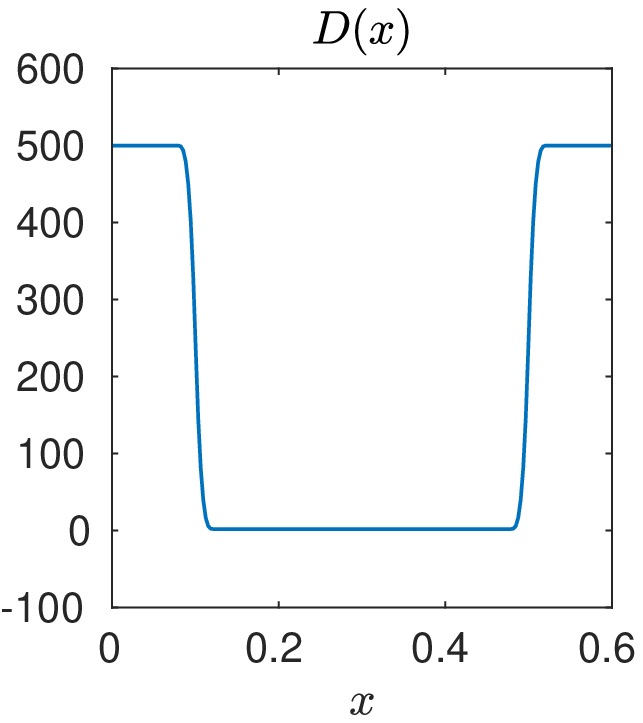}
  \label{fig:D}~}
\subfloat[Maxwellian]
{\includegraphics[width=0.305\linewidth]{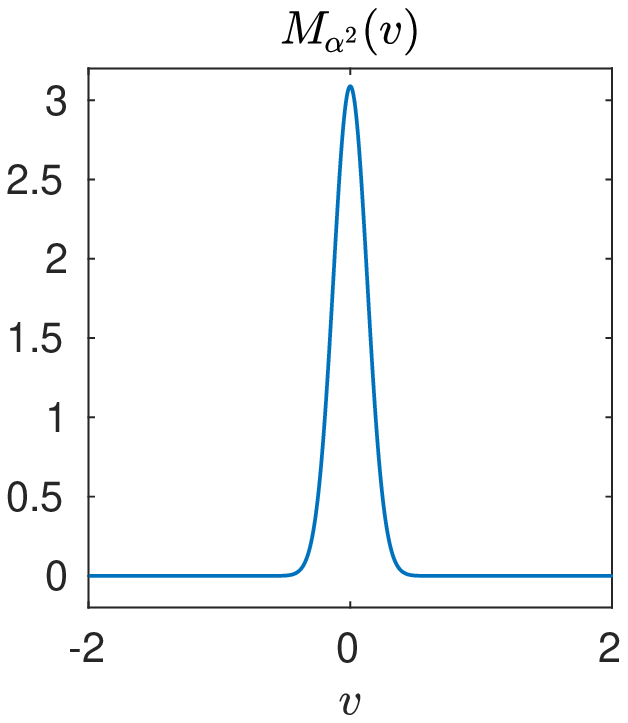}
  \label{fig:M}~}
\subfloat[Collision frequency]
{\includegraphics[width=0.344\linewidth]{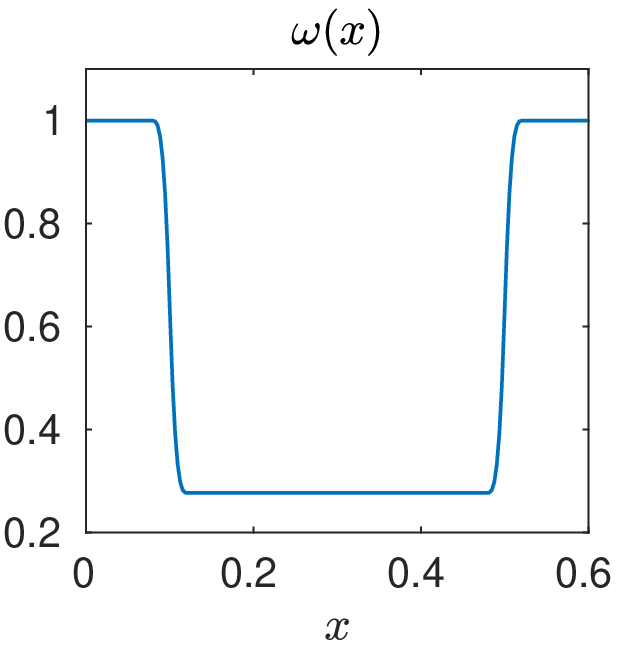}
  \label{fig:nu}}
\caption{The scaled doping profile, Maxwellian, and collision frequency for the standard $n^+$--$n$--$n^+$ diode problem. }
\label{fig:D_M_nu}
\end{figure}

\begin{figure}[h]
   \captionsetup[subfigure]{justification=centering}
\subfloat[Electron concentration]
{\includegraphics[width=0.325\linewidth]{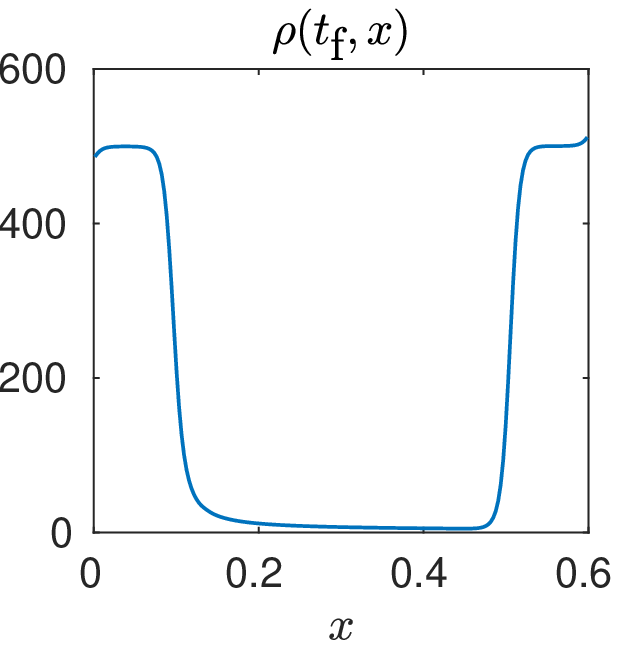}
  \label{fig:rho}~}
\subfloat[Electric field]
{\includegraphics[width=0.31\linewidth]{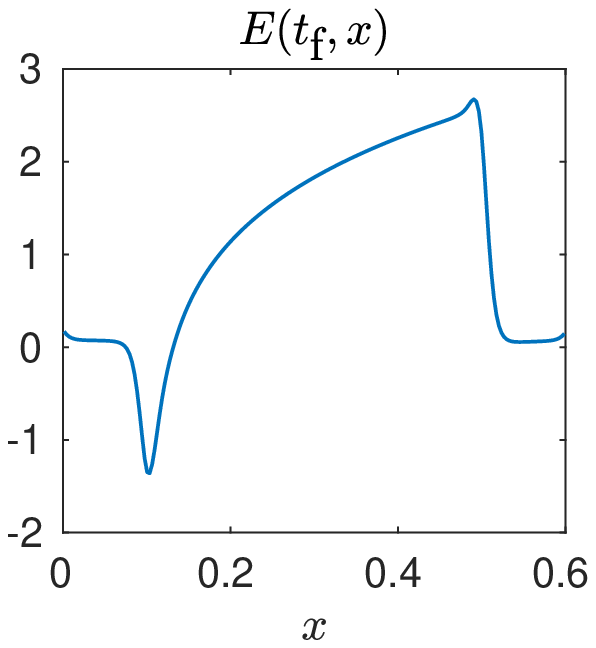}
  \label{fig:E}~}
\subfloat[Electron distribution]
{\includegraphics[width=0.38\linewidth]{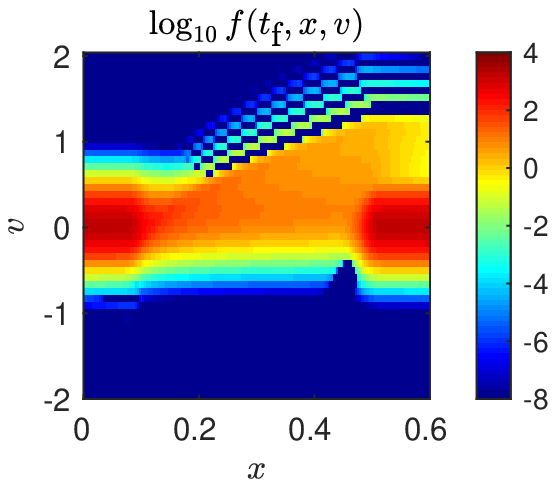}
  \label{fig:f}}
\caption{The electron concentration, electric field, and electron distribution at the final time $t_\textup{f}=0.5$ for the standard $n^+$--$n$--$n^+$ diode problem with scaled collision frequency given in \eqref{eq:nu_standard_scaled}. Here the electron distribution $f$ is plotted in Figure~\ref{fig:f} in logarithmic scale, and the magnitude of the observed oscillations at positive velocities is smaller than 0.1\% of the maximum value of $f$. These oscillations are stable artifacts that can be mitigated or removed by refining the velocity discretization.}
\label{fig:rho_E_f}
\end{figure}

Figure~\ref{fig:D_M_nu} shows the scaled doping profile $D(x)$ with smooth transitions, Maxwellian $M_{\alpha^2}(v)$, and collision frequency $\omega(x)$ in this silicon diode problem.
Figure~\ref{fig:rho_E_f} illustrate the electron concentration $\rho$, the electric field $E$, and the full electron distribution $f$ at the final, steady-state time $t_\textup{f}=0.5$.
As expected, all tested solvers give identical results up to the set numerical tolerance, and these results agree with the those reported in other places in the literatures, such as \cite{Carrillo2000iza,Cercignani2000domain}.


Table~\ref{table:nnn_iter_time} reports the iteration counts and computation time for each of the iterative solvers.
The system is away from the drift-diffusion limit with the given collision frequency.
Hence the results in Table~\ref{table:nnn_iter_time} are similar to the ones in Table~\ref{table:single_scale_iter_time_1} and the drift-diffusion synthetic acceleration does not accelerate the convergence of the iterative solvers.
The type-\Rmnum{1} solvers still outperform the type-\Rmnum{2} solvers for this test.

Here the implicit time step $\dt=\dx$ is roughly 35x larger than the standard explicit and IMEX-AP time steps $\dt\leq\delta {\dx}/{v_{\max}}$.
We observe in Table~\ref{table:nnn_eff} that the implicit scheme generally gives $\bigchi>1$, except when the accelerated type-\Rmnum{2} AA/GMRES solver is used.

\begin{table}[h]
\centering
\begin{tabular}{|| c || c | r | r | r || c || c | r | r | r | r ||}
\toprule
\multirow{2}{*}{Solver} & \multirow{2}{*}{SA} &
\multicolumn{2}{c|}{Iteration} & {Total} &\multirow{2}{*}{Solver} & \multirow{2}{*}{SA} &
\multicolumn{3}{c|}{Iteration} & {Total}\\
\cline{3-4} \cline{8-10}
 & & {FP} & {SW} & {Time} &  & & {FP} & {LS} & {SW} & {Time} \\
\hline
 {Type-\Rmnum{1}}  
 & N & 3.8 & 1.4 &  7.6 & {Type-\Rmnum{2}}  & N & 2.6 & 4.8 & 1.4 & 28.6 \\
 \cline{2-5}  \cline{7-11}
 {PI} 
 & Y & 4.6 & 1.6 & 10.1 & {AA/PI}  & Y & 2.6 & 4.5 & 1.4 & 29.6 \\
\hline
 {Type-\Rmnum{1}}  
 & N & 2.6 & 1.4 &  7.3 & {Type-\Rmnum{2}}  & N & 2.8 & 2.2 & 3.7 & 49.8 \\
 \cline{2-5}  \cline{7-11}
 {AA} 
 & Y & 2.8 & 1.4 &  8.2 & {AA/GMRES}  & Y & 6.0 & 3.5 & 3.6 & 110.7 \\
\bottomrule
\end{tabular}
\caption{Standard $n^+$--$n$--$n^+$ diode problem with $\omega(x)$ given in \eqref{eq:nu_standard_scaled}, $\delta=\epsilon=0.056$, and $\dt = \dx$. Iteration counts and total computation time (sec) for the compared solvers and their accelerated variants.}
\label{table:nnn_iter_time}
\end{table}

\begin{table}[h]
	\centering
	\begin{tabular}{|| c || c | c | c | c || c || c | c | c | c ||}
		\toprule
		\multicolumn{10}{||c||}{$1$ implicit step $\sim$ $35$ IMEX-AP steps $\sim$ $35$ explicit steps}\\
		\hline
		Solver & SA & {Iter$_{\text{SW}}$} & {$\bigchi_{\text{IMEX}}$} & {$\bigchi_{\text{EX}}$} & Solver & SA & {Iter$_{\text{SW}}$} & {$\bigchi_{\text{IMEX}}$} & {$\bigchi_{\text{EX}}$}\\
		\hline
		{Type-\Rmnum{1}}  	& N & 5.3 & 6.58 & 6.58 & {Type-\Rmnum{2}}  & N & 17.5 & 2.00 &  2.00\\
		\cline{2-5}  \cline{7-10}
		{PI}  				& Y & 7.4 & 4.76 & 4.76 & {AA/PI} 			& Y & 16.4 & 2.14 &  2.14 \\
		\hline
		{Type-\Rmnum{1}}   	& N & 3.6 & 9.62 & 9.62 & {Type-\Rmnum{2}} & N & 22.8  & 1.54 &  1.54 \\
		\cline{2-5}  \cline{7-10}
		{AA} 				& Y & 3.9 & 8.93 & 8.93 & {AA/GMRES}  		& Y & 75.6 & 0.46 &  0.46 \\
		\bottomrule
	\end{tabular}
	\caption{Standard $n^+$--$n$--$n^+$ diode problem with $\omega(x)$ given in \eqref{eq:nu_standard_scaled}, $\delta=\epsilon=0.056$, and $\dt = \dx$.  The values of the efficiency indicator $\bigchi$ \eqref{eq:chi_def} and the total number of sweeping iterations in an implicit step (Iter$_{\text{SW}}$) are reported. This serves as a formal efficiency comparison between the proposed implicit scheme and the standard explicit and IMEX-AP schemes.}
	\label{table:nnn_eff}
\end{table}

We next consider another multiscale problem with stronger variation in the collision frequency.
Specifically, the scaled collision frequency considered in this problem is
\begin{equation}\label{eq:multiscale_nu_scaled}
\omega(x) = 
\begin{cases}
0.01, &  x \in [0.1, 0.5] \\
1, &  \textup{otherwise}
\end{cases}
\end{equation}
with $\omega_0=5.114 \times 10^{14}  \si{s^{-1}}$ and $\delta=\epsilon = \frac{v_0 }{x_0 \omega_0}=0.002$.

The iteration counts and computation time for this problem are reported in Table~\ref{table:multi_scale_iter_time}. 
These results show that for this multiscale problem, synthetic acceleration with the drift-diffusion model speeds up the convergence of most iterative solvers by roughly a factor of two.
However, we observe diverging residuals (denoted as \DIV) when using the accelerated type-\Rmnum{1} PI scheme, which indicates that applying the drift-diffusion based synthetic acceleration on multiscale problems may lead to unstable schemes that give divergent results.
This observation is related to a known deficiency of diffusion synthetic acceleration (DSA).
Specifically, it has been reported in \cite{Gelbard-Hageman-1969} and analyzed in \cite{Reed-1971} that DSA becomes unstable and gives divergent results when applied to highly collisional problems.
To guarantee convergence, the spatial discretization of the diffusion equation has to be consistent to the discretization of the original transport equation.
We refer the reader to \cite{Adams-Larsen-2002} and \cite{Larsen-Morel-2010} for a complete discussion.

Finally, we note that the standard explicit and IMEX-AP time steps both satisfy $\dt\leq\delta\dx/v_{\max}$, which are 1000x smaller than the implicit time step $\dt=\dx$.
As shown in Table~\ref{table:multi_scale_eff}, the efficient indicator $\bigchi>1$ for all iterative solvers that lead ti a convergent implicit scheme.
Among these iterative solvers, the most efficient one is the accelerated type-\Rmnum{1} AA solver.

\begin{table}[h]
\centering
\begin{tabular}{|| c || c | r | c | r || c || c | c | r | c | r ||}
\toprule
\multirow{2}{*}{Solver} & \multirow{2}{*}{SA} &
\multicolumn{2}{c|}{Iteration} & {Total} &\multirow{2}{*}{Solver} & \multirow{2}{*}{SA} &
\multicolumn{3}{c|}{Iteration} & {Total}\\
\cline{3-4} \cline{8-10}
 & & \multicolumn{1}{c|}{FP} & {SW} & {Time} &  & & {FP} & \multicolumn{1}{c|}{LS} & {SW} & {Time} \\
\hline
 {Type-\Rmnum{1}}  	& N & 79.4& 1.3 &125.7& {Type-\Rmnum{2}}  	& N & 1.5 &107.2& 1.3 &395.0\\
 \cline{2-5}  \cline{7-11}
 {PI}  				& Y & \multicolumn{1}{c|}{\DIV} & \DIV & \multicolumn{1}{c||}{\DIV} & {AA/PI} 				& Y & 1.7  & 42.4 & 1.3  & 184.7\\
\hline
 {Type-\Rmnum{1}}   	& N & 14.2 & 1.3 &  26.0 & {Type-\Rmnum{2}}  	& N & 1.9 & 30.0 & 3.0 & 207.8 \\
 \cline{2-5}  \cline{7-11}
 {AA} 				& Y & 3.9 & 1.1 & 10.0& {AA/GMRES}  			& Y & 2.2 & 9.9 & 2.8 &92.0 \\
\bottomrule
\end{tabular}
\caption{Multiscale problem with $\omega(x)$ given in \eqref{eq:multiscale_nu_scaled}, $\delta=\epsilon=0.002$, and $\dt = \dx$. Iteration counts and total computation time  (sec) for the compared solvers and their accelerated variants. Here {\DIV} denotes the case that diverging residual is observed.}
\label{table:multi_scale_iter_time}
\end{table}

\begin{table}[h]
	\centering
	\begin{tabular}{|| c || c | r | r | r || c || c | r | r | r ||}
		\toprule
		\multicolumn{10}{||c||}{$1$ implicit step $\sim$ $1000$ IMEX-AP steps $\sim$ $1000$ explicit steps}\\
		\hline
		Solver & SA & {Iter$_{\text{SW}}$} & {$\bigchi_{\text{IMEX}}$} & \multicolumn{1}{c||}{$\bigchi_{\text{EX}}$} & Solver & SA & {Iter$_{\text{SW}}$} & {$\bigchi_{\text{IMEX}}$} & \multicolumn{1}{c||}{$\bigchi_{\text{EX}}$}\\
		\hline
		{Type-\Rmnum{1}}  	& N & 103.2 & 9.69 & 9.69 & {Type-\Rmnum{2}}  & N & 209.0 & 4.78 & 4.78 \\
		\cline{2-5}  \cline{7-10}
		{PI}  				& Y & \multicolumn{1}{c|}{\DIV} & \multicolumn{1}{c|}{---} & \multicolumn{1}{c||}{---} & {AA/PI} 			& Y & 93.7 &10.67 & 10.67 \\
		\hline
		{Type-\Rmnum{1}}   	& N & 18.5 &54.05 &54.05 & {Type-\Rmnum{2}}  & N & 171.0 & 5.85 &  5.85 \\
		\cline{2-5}  \cline{7-10}
		{AA} 				& Y & 4.3 &232.56&232.56& {AA/GMRES}  		& Y & 61.0 &16.39 & 16.39 \\
		\bottomrule
	\end{tabular}
	\caption{Multiscale problem with $\omega(x)$ given in \eqref{eq:multiscale_nu_scaled}, $\delta=\epsilon=0.002$, and $\dt = \dx$.  The values of the efficiency indicator $\bigchi$ \eqref{eq:chi_def} and the total number of sweeping iterations in an implicit step (Iter$_{\text{SW}}$) are reported. This serves as a formal efficiency comparison between the proposed implicit scheme and the standard explicit and IMEX-AP schemes.}
	\label{table:multi_scale_eff}
\end{table}
\section{Conclusions and discussion}
\label{sec:conclusion}
We have proposed a fully implicit numerical scheme for solving Boltzmann-Poisson systems with an approximate collision operator that describes linear relaxation.  At each implicit time step, the updated solution comes from a nonlinear fixed-point problem.
We have formulated Boltzmann-Poisson systems as two types of fixed-point problems: the type-\Rmnum{1} problems that are solved in a single strongly coupled iterative loop and the type-\Rmnum{2} problems that require two nested iterative loops.
We have applied Picard iteration and Anderson acceleration to solve the type-\Rmnum{1} problems.
For the type-\Rmnum{2} problems, the outer loop of the nested iterative procedure is performed by Anderson acceleration while the inner loop uses either Picard iteration or GMRES.
The performance of these iterative solvers and their synthetically accelerated (SA) variants is compared on several scaled versions of a standard silicon diode problem.
Numerical results show that (i) solving the type-\Rmnum{1} fixed-point problems requires fewer iterations than solving the type-\Rmnum{2} problems and (ii) Anderson acceleration is more efficient than Picard iteration on type-\Rmnum{1} problems, in terms of both iteration counts and computation time.
These results also confirm that SA schemes with a drift-diffusion model converge faster than the standard iterative solvers when the system is near the drift-diffusion limit (when collision frequency is large). 
For systems away from this limit, there is no observed advantage in using these accelerated schemes.
To address this issue, some potential approaches to be investigated in the future include (i) modifying the drift-diffusion SA schemes to incorporate the boundary conditions as proposed in \cite{valougeorgis1988boundary} for a neutron transport problem and (ii) deriving SA schemes based on low-order/low-cost approximations other than the drift-diffusion equation.
For (ii), possible candidates of such approximation include the $S_2$-SA scheme considered in \cite{lorence1989s,Bruss-Morel-Ragusa-2014} and the more general, two-level multigrid algorithms as the one found in \cite[Section~6.2]{atkinson_1997}.

Another potential future work is to apply the hybrid schemes \cite{Crockatt-2017, Hauck-McClarren-2013} proposed for linear transport equations.
These schemes decompose the transport solution into collisional and non-collisional components.
At each time step, the hybrid schemes use cheap, low-resolution approximations for the collisional component and reserve expensive, high resolution approximations for the the non-collisional component.
Since the primary difficulty on solving the Boltzmann-Poisson system is the nonlinear coupling between the electric field and collision term, we expect that the hybrid approach would significantly lower the computation cost while producing quality solutions that are comparable to the ones given by an uniformly high-resolution solver.

\bibliographystyle{plain}
\bibliography{reference}

\end{document}